\documentclass[apj,twocolumn,twocolappendix,appendixfloats]{openjournal}

\usepackage{lipsum}

\usepackage{xcolor}
\usepackage{textgreek}
\usepackage[utf8]{inputenc}
\usepackage[english]{babel}

\usepackage{hyperref}
\hypersetup{
    unicode, 
    colorlinks=true,
    linkcolor=linkcolor,
    citecolor=linkcolor,
    filecolor=linkcolor,
    urlcolor=linkcolor,
}
\usepackage{color,colortbl}
\definecolor{linkcolor}{rgb}{0.0,0.3,0.5}
\usepackage{tensind}
\tensordelimiter{?}
\DeclareGraphicsExtensions{.bmp,.png,.jpg,.pdf}
\usepackage{verbatim}
\usepackage[normalem]{ulem}
\usepackage{orcidlink}
\usepackage{soul}
\usepackage{amsmath}
\usepackage[caption=false]{subfig}
\usepackage{multirow}
\usepackage{graphicx,float}
\graphicspath{ {./figs/} }
\usepackage{makecell}

\begin{document}
\title{On the relation between magnetic field strength and gas density in the interstellar medium. II. Density uncertainties and diffuse gas constraints}

\author{David Whitworth$^{1}$\orcidlink{0000-0001-9956-0785}\email{david.whitworth@ens-lyon.fr},
Amit Seta$^{2}$\orcidlink{0000-0001-9708-0286},
Ralph E. Pudritz$^{3}$\orcidlink{0000-0000-0000-0000},
Mordecai-Mark Mac Low$^{4}$\orcidlink{0000-0003-0064-4060},
Juan~D. Soler$^{5}$\orcidlink{0000-0002-0294-4465},
Aina Palau$^{6}$\orcidlink{0000-0002-9569-9234},
Ralf S.~Klessen$^{7,8}$\orcidlink{0000-0002-0560-3172}}

\affiliation{
$^{1}$ENS de Lyon, CRAL UMR5574, Universite Claude Bernard Lyon 1, CNRS, Lyon 69007, France \\
$^{2}$Research School of Astronomy and Astrophysics, Australian National University, Canberra, ACT 2611, Australia \\
$^{3}$Department of Physics and Astronomy, McMaster University, 1280 Main Street West, Hamilton, ON, L8S4K1, Canada \\
$^{4}$Department of Astrophysics, American Museum of Natural History, 200 Central Park West, New York, NY 10024, USA \\
$^{5}$Department of Astrophysics, T\"urkenschanzstra{\ss}e 17 (Sternwarte), 1180 Wien, Austria \\
$^{6}$Universidad Nacional Aut\'onoma de M\'exico, Instituto de Radioastronom\'ia y Astrof\'isica, Antigua Carretera a P\'atzcuaro 8701, Ex-Hda. San Jos\'e de la Huerta, 58089, Morelia, Michoac\'an, M\'exico \\
$^{7}$Universit\"{a}t Heidelberg, Zentrum f\"{u}r Astronomie, Institut f\"{u}r Theoretische Astrophysik, Albert-Ueberle-Str.\ 2, 69120 Heidelberg, Germany \\
$^{8}$Universit\"{a}t Heidelberg, Interdisziplin\"{a}res Zentrum f\"{u}r Wissenschaftliches Rechnen, Im Neuenheimer Feld 225, 69120 Heidelberg, Germany}

\begin{abstract}
The relationship between magnetic field strength and gas density is essential to understand the interstellar medium and star formation. Zeeman measurements in dense atomic and molecular gas phases have traditionally been used to directly probe magnetic field strengths in the Milky Way. This allowed derivation of a relationship between magnetic field strength $B$ and gas number density $n$. We recently generalized this relation as a two-part power-law with non-zero slopes and a transition density given as $B/B_0 \propto (n/n_0)^{\alpha_1}$ for $n \le n_0$ and $(n/n_0)^{\alpha_2}$ for $n > n_0$. Here, we extend our previous hierarchical Bayesian framework by incorporating a large body of pulsar observations that probe the diffuse interstellar medium and explicitly modelling density uncertainties through a log-density intrinsic scatter $R_n$ applied to all densities. We also account for magnetic field geometry and measurement uncertainties through a magnetic hyperparameter to estimate $B$. This results in a stronger constraint on the diffuse gas part of the $B$--$n$ relation. Our results confirm a non-zero exponent in the diffuse gas and a broad transition density with our best model and data set yielding maximum a posteriori results of $\alpha_1 = 0.20^{+0.01}_{-0.03}$, $\alpha_2 = 0.65^{+0.06}_{-0.07}$, $n_0 = 3330^{+270}_{-3220}\,\text{cm}^{-3}$, and $B_0 = 10.11^{+0.71}_{-5.97}\,\mu\text{G}$.
\end{abstract}


\maketitle

\section{Introduction}
\label{sec:intro}

Magnetic fields are a dynamically important component of the interstellar media (ISM) of galaxies, affecting processes as diverse as cosmic ray propagation, pressure support in diffuse gas, and star formation in the denser phases. In the latter case, although the details remain uncertain \citep[see for example][]{Hennebelle2019,Pattle2022} theoretical and computational work suggest that magnetic fields can support molecular clouds and filaments against large scale collapse, reducing star formation rates \citep{Mestel1956, Nakano1973,Mouschovias1976, Fiege_Pudritz2000a,Robinson2023,Pillsworth2025a,Bogue2026}. The density at which the dynamical impact of magnetic fields changes may also be particularly important during the transition from warm to cold, star-forming gas \citep{McGuiness2025}. This, in turn, can affect the life cycle of giant molecular clouds, potentially changing how much molecular gas forms from the more diffuse ISM \citep{Whitworth2023}. Understanding this transition from diffuse to dense gas and where it occurs in the ISM is key to our understanding of molecular cloud formation and evolution, which in turn is important for explaining star formation and galaxy evolution.

\defcitealias{Crutcher2010}{C10}
\defcitealias{Whitworth2025}{Paper~I}
Given that physical processes in the diffuse ISM differ from those in the molecular, self-gravitating gas in which stars form, it is natural to investigate a two-part broken power law relation between magnetic field strength and gas density in the diffuse vs.\ denser gas. This gives rise to a general $B$--$n$ relation of the form 
\citetext{\citealt{Crutcher2010}, hereafter \citetalias{Crutcher2010}, \citealt{Jiang2020}, \citealt{Whitworth2025}, hereafter \citetalias{Whitworth2025}}
\begin{equation}
\label{power_law}
    |\mathbf{B}| = B_0
    \begin{cases}
      (n/n_0)^{\alpha_1} & \mbox{if } n \leq n_0,\\
      (n/n_0)^{\alpha_2} & \mbox{if } n > n_0,
    \end{cases}  
\end{equation}
where $B_0$ and $n_0$ are the normalisation parameters, and $\alpha_1$, $\alpha_2$ are the power-law indices on either side of the break density, $n_0$. The value of the critical density $n_0$ characterizing the transition between these two regimes depends upon the power law indices. Thus, the first Bayesian analysis of the Zeeman data by \citetalias{Crutcher2010} \citep[also see][]{Jiang2020} gave critical densities of $\sim 300\,\mathrm{cm}^{-3}$, while assuming that $\alpha_1 = 0$. 

In more recent work, no restrictive assumptions are made about the power-law indices \citepalias{Whitworth2025}, resulting in a wider range of critical densities; $4\,000^{+13\,000}_{-3\,000}\,\mathrm{cm}^{-3}$ and $\alpha_1 = 0.15^{+0.06}_{-0.09}$. The results of this latter, more general Bayesian analysis are in agreement with magnetohydrodynamical simulations of dwarf galaxies \citep{Whitworth2025,McGuiness2025}. \citet{McGuiness2025} showed that the field is dominant across the warm to cold phase transition regime of $1 < n \lesssim 1\,000\,\mathrm{cm}^{-3}$.

Different tracers of the magnetic field probe distinct phases of the ISM. Thus, assembling a unified $B$--$n$ relation across the entire ISM remains a challenge \citep{Beck2015,Ferriere2020,Seta2022,Gent2023}. As an example, polarized dust emission can be analysed with the Davis-Chandresekar-Fermi (DCF) method to estimate the magnetic field strength in dense molecular clouds, Zeeman measurements typically estimate the field strength in mid- to high-column density regions, while pulsar rotation measures probe the more diffuse, ionized medium. Each of these techniques measures different components or projections of the magnetic field: the line-of-sight (LOS) component via Zeeman splitting and rotation measure, and the plane-of-sky (POS) component via the DCF method. \citetalias{Whitworth2025} showed that a two-part power-law arises when including data analysed with the DCF method. 

However, care must be taken when comparing these data because they measure different projections of the magnetic field.

Furthermore, in \citetalias{Whitworth2025} the DCF fit appears elevated relative to Zeeman data due to known systematic over-predictions in DCF by factors of $\sim 4$--$6$ arising from line-of-sight averaging and turbulence assumptions \citep[see e.g.,][]{Chen2022,Pattle2022}. Therefore, this paper focuses exclusively on LOS $B$ measurements (Zeeman and pulsars) to maintain methodological and projection consistency.

There are at least two major considerations in constraining the observationally derived $B$--$n$ relationship. The first is to secure broad and deep observational data sets. The diffuse ISM and molecular gas taken together cover many decades in density over which accurate measurement of field strengths is, at the very least, challenging. Although the observational data used in \citetalias{Whitworth2025} covered a wide range of density and field strengths, it still lacked measurements of magnetic fields at very low densities in the diffuse gas. In this paper, we update our analysis by including in our sample values of the average LOS $B$ for over 200 pulsars in the diffuse ISM \citep{Seta2025}.

The second consideration is that both variables have substantial measurement uncertainties. The volume number density $n$ cannot be measured directly, so it must be inferred from variables such as temperature, column density, and model-dependent assumptions about cloud geometry and excitation conditions \citep{Pineda2010,Leroy2017}. The new pulsar-based LOS $B$ estimates, derived from rotation and dispersion measures \citep[e.g.][]{Seta2025, Dhakal2025}, are similarly sensitive to LOS integration effects. In contrast, Zeeman splitting estimates \citep[e.g.][]{Heiles2005,Crutcher2012} provide more localized constraints but are susceptible to measurement noise and LOS ambiguity \citepalias{Crutcher2010}. Therefore, accurate inferences about the $B$--$n$ scaling require careful propagation of uncertainties in both quantities. In this paper, we provide a more sophisticated treatment of the errors than in \citetalias{Whitworth2025} by running a more general Bayesian hierarchical treatment using an extended set of hyperparameters.

To illustrate the importance of obtaining better and more extensive measurements and error estimates in modelling the $ B$--$n $ relationship, we note that a different kind of scaling has been derived using the pulsar data. It is based upon the comparison of turbulent kinetic and magnetic energy densities \citep{Seta2025}. One characteristic of this model is that it overcomes the necessity of reconstructing the volumetric density, as the errors in turbulent velocities strongly outweigh the errors in number density. The result is a single, linear magnetic to turbulent kinetic energy density relation. However, this does not help our understanding of the transition between diffuse ISM and dense molecular clouds, and it becomes difficult to isolate the role that the magnetic field plays here.

One thing that has yet to be considered in any analysis of the $B$--$n$ relationship is the Zeeman observational sensitivity threshold. All observations suffer from constraints set by the signal-to-noise ratio (S/N). \cite{Troland2008} estimate that for their observations, an S/N of $\sim 2.4$ can marginally detect a field strength of $\sim 2.5\,\mu$G. In practice, this means that fields below $\sim 2.5\,\mu$G are increasingly difficult to distinguish from noise and may be reported as non-detections. This effect was not explicitly accounted for in \citetalias{Crutcher2010} and subsequent work. This could lead to a bias in the fit. Pulsar-based magnetic field estimates are also subject to observational selection effects, but these are expressed as limitations on rotation and dispersion measures rather than a direct field-strength detection threshold; consequently, they are not affected in the same way as Zeeman non-detections.

\citetalias{Whitworth2025} implemented a Bayesian analysis for multiple numerical simulations in order to compare simulations with observations within the same framework. We examined 19 numerical simulations, ranging from protostellar discs to Milky Way-like galaxies, and focused on the physical factors that drive the $B$--$n$ relation. In all cases, we found that a two-part broken power law provided the best fit to the numerical data, although the values of the parameters varied. We found that numerical limitations, such as resolution---full galaxy models fail to capture the smallest scales and to resolve the dynamo in diffuse gas---and time---box models may not evolve for long enough to overcome the initial conditions of the field---affect the values of the parameters. 

Several recent studies have examined the relationship in other ways and show results in broad qualitative agreement with ours. \citet{Zhao2024_b} analysed a magnetohydrodynamical giant molecular cloud model, finding a gradual decrease in the importance of the field as density increases, suggesting a continuous, curved relationship with no single well-defined break. \cite{Cao2023} simulated dense core evolution, showing a temporal evolution. \cite{Seta2022} showed in multi-phase simulations that the relationship is complex and cannot be described by a single power-law. \cite{Konstantinou2024} argue that, based on a Milky Way-like galaxy simulation, the relationship is not universal and displays variation over time. In a suite of cosmological zoom-in simulations, \citet{Ponnada2022} found that the relationship does not vary over a large range of densities, showing no distinct break density in any of their models.

In this paper, we build upon \citetalias{Whitworth2025} by including LOS magnetic field measurements deduced from pulsar data from \citet{Seta2025}, which includes densities at least two orders of magnitude lower than in \citetalias{Crutcher2010}, reaching $n <10^{-2}$\,cm$^{-3}$. Also, we implement a more general hierarchical Bayesian analysis of the errors by exploring four models with more extensive parameter sets than used in our previous paper. In Section \ref{sec:method}, we introduce the new extended data set and our four models. Section \ref{sec:results} shows the results from these extended models and data sets. Section \ref{sec:discussion} discusses the results and how the different approaches affect the relationship. Finally, Section \ref{sec:conclusions} summarises our key findings. We also include four appendices, Appendix~\ref{Append:code}, correcting an error in \citetalias{Whitworth2025}; Appendix~\ref{ap:plots}, presenting the corner plots for our Bayesian models; Appendix~\ref{ap:sensitivity}, addresses the sensitivity limit of the Zeeman data; and Appendix~\ref{app:single_slope}, presents a single-slope model.

\section{Methods}
\label{sec:method}

\subsection{Previous treatments of errors}
\label{subsec:errors}
\citetalias{Crutcher2010} used reported errors on $B$. They described the uncertainty in the observed density value $n_\mathrm{obs}$ as a fixed error of a factor of two. \citet{Tritsis2015} and \citet{Jiang2020} reanalysed the data using different approaches and showed that a factor of two is likely too low. \citet{Tritsis2015} used a more relaxed assumption and a least squares fit approach, while \citet{Jiang2020} used a Bayesian approach allowing the error to be a free parameter reporting possible values ranging from $7.7 - 44.1$, with $9.3$ being the most likely. \citetalias{Whitworth2025} also reanalysed the data using three fixed values for the error, $2, 5$, and $9$, reporting their best result with a value of $5$. However, this assumes that each data point has the same error, which is highly unlikely due to the different environments, telescopes, and projection effects at play. In principle, the most rigorous approach would involve assigning an individual measurement uncertainty to each data point and propagating these through the Bayesian framework. Unfortunately, such detailed error estimates are rarely reported, leaving us to rely on an intrinsic scatter hyperparameter $R_n$ as an approximation for the error of $n$.

There have been more recent Zeeman surveys of magnetic field strengths, but of these, only \citet{Hwang2024} report estimates of the number density $n$, with the rest only providing column densities \citep{Thompson2019,Nakamura2019,Koley2021,Ching2022,Nowotka2025}. Inferring three-dimensional densities from two-dimensional, LOS observations is challenging. The original Zeeman data (\citealt{Crutcher2010}) densities were derived using tracer-specific methods (e.g.\ H\,{\sc i} spin temperature with interstellar-pressure assumptions; \citealt{Heiles2005,Crutcher2010}), without requiring detailed prior constraints on cloud geometry, morphology, or precise distances. When equivalent tracer-level density information is unavailable, volume densities may instead be estimated from column-density data using additional structural information \citep[e.g.][]{Crutcher2012}. Whilst the combination of line emission observations with reconstructed three-dimensional dust mapping \citep{Edenhofer2024} can help identify the LOS distribution of the gas \citep[see, for example,][]{Zucker2021,Angarita2024,Soler2025}, it demands high-resolution density reconstructions both along the LOS and across the plane of the sky (with three-dimensional dust maps matching the angular resolution and pixel scales of the line emission data), plus prior assumptions about cloud/filament morphology.

\subsection{Extended data sets: pulsars}

In \citetalias{Whitworth2025}, we used the original 137 Zeeman data points from \citetalias{Crutcher2010}, which targeted cold neutral H~{\sc i} with $10 \lesssim n \lesssim 300 \mbox{ cm}^{-3}$ as well as higher-density OH and CN measurements in molecular clouds. In this paper, we further include the additional six Zeeman measurements from \citet{Hwang2024}. We exclude DCF data here due to its plane-of-sky geometry and factor of $\sim 6$ over-prediction relative to LOS methods \citep[e.g.][]{Pattle2022}, which complicates joint Bayesian modelling without custom likelihood adjustments for these systematics; our LOS-only focus enables robust extension to pulsars probing low-density gas \citep{Seta2025}. A recent study by \citet{Nowotka2025} presents new FAST Zeeman measurements and examines LOS averaged data. While their analysis of column-density scaling in the diffuse ISM (e.g., their Figure 10) offers a useful comparison for low-density trends, we do not explicitly incorporate their dataset into our framework. 

The pulsar measurements from \citet{Seta2025} that we use in this paper include a total of 212 points that probe down to densities of $\sim 10^{-2} \mbox{ cm}^{-3}$. This is critically important because these low density points provide considerable leverage on the diffuse part of the $B$--$n$ relation not otherwise obtainable. These observations derive average electron density from the ratio of the dispersion measure to pulsar distance, and the line-of-sight component of the magnetic field from the ratio between the rotation and dispersion measures, assuming uncorrelated thermal electron density and magnetic fields along the line of sight \citep{SetaF2021}. These methods have long been used to study Galactic magnetic fields in low-density gas \citep{Hewish1968,Smith1968,Manchester1974, Han2017}.

The Zeeman and pulsar samples were obtained with different observational setups, reduction pipelines, and calibration procedures, so the compiled catalogue is not homogeneous. This introduces systematic biases due to differences in angular resolution, sensitivity, selection criteria and calibration techniques, and the use of gas densities that may not always probe the exact same gas volume as the magnetic-field measurement in the Zeeman data sets. Combining data obtained at different spatial scales and with different tracers can alter the inferred $B$--$n$ relation and/or introduce additional statistical fluctuations. Nevertheless, combining these data sets remains valuable, as it extends the dynamic density range far beyond what either sample alone can provide. To check the extent of the resulting biases, we analyse both Zeeman-only data sets and combined Zeeman+pulsar data sets, comparing the resulting fits. For full details of the observations and reduction procedures for the observational data, we refer the reader to the original publications.

An important consideration in any analysis combining multiple surveys is probe-specific systematics. Pulsar rotation and dispersion measure estimates trace LOS-averaged, ionized-gas-weighted quantities and can be affected by field reversals and path-length averaging, which may reduce inferred $|B_\parallel|$ at fixed effective density. In contrast, Zeeman measurements are object- and tracer-dependent. H\,{\sc i}, OH, and CN probe different physical regimes and rely on different assumptions when converting observables into representative densities. Additional effects arise from beam averaging, excitation and abundance assumptions, and LOS structure. Together, these can introduce both additional scatter and tracer-dependent offsets in the inferred $B$--$n$ relation. In our hierarchical framework, we implement specific hyperparameters (Section~\ref{sec:model}) that can factor in part of this residual variance from unresolved measurement and systematic effects for the power-law fitting analysis, but they likely do not fully remove tracer-specific biases. We therefore assess robustness by comparing Zeeman-only and Zeeman+pulsar datasets, with and without the outlier (DS1--DS4), and across model variants with fixed versus inferred density scatter (Model~A--D). For detailed survey-level observational and reduction systematics, we refer readers to the original data papers: pulsars \citep{Manchester2005} and Zeeman \citep{Crutcher1999,Heiles2004,Heiles2005,Falgarone2008,Troland2008}.

\begin{figure}
    \centering
    \includegraphics[scale=0.35]{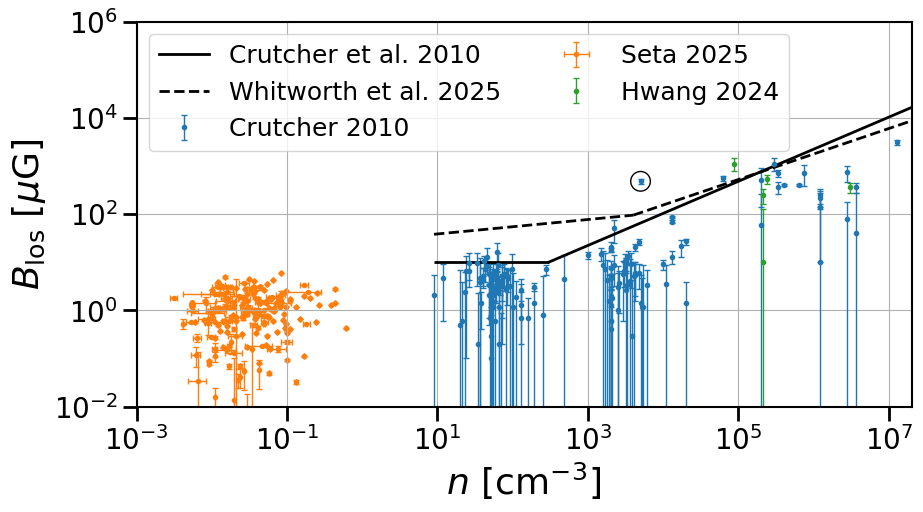}
    \caption{Full data set including Zeeman data from \protect \citetalias{Crutcher2010} and \protect \cite{Hwang2024} and pulsar data from \protect \cite{Seta2025}. Lines show \protect\citetalias{Crutcher2010} upper-limit power law {\em (solid line)} and the \protect \citetalias{Whitworth2025} MAP result {\em (dashed line)}, which, as discussed in Appendix~\ref{Append:code}, had too high a value of $B_0$. (Note that Fig.~1 of \protect \citealt{Seta2025} used the preliminary result from an early preprint version of \protect\citetalias{Whitworth2025}.) The outlier in the Zeeman data is circled for identification, see main text for further details \protect\citepalias{Crutcher2010}.}
    \label{fig:data}
\end{figure}

The combined data expands the number density range beyond that considered in \citetalias{Whitworth2025} to $ 10^{-2}$ cm$^{-3} < n < 10^7 \mbox{ cm}^{-3}$, as well as increase the number of data points from 137 to 347. In Figure \ref{fig:data} we show the entire extended data set. One point circled in the Zeeman data lies above the bulk of the data and the original relation in \citetalias{Crutcher2010}.\footnote{\citetalias{Crutcher2010} note that this point corresponds to the giant molecular cloud Sgr B2 North, suggesting it likely has significant errors on $n$. This interpretation is supported by \citet{Jiang2020}, whose Bayesian Model C, using a framework similar to \citetalias{Crutcher2010}, showed little variation under the assumption of this outlier. However, when they modified the Bayesian setup, the outlier became more significant.} The inclusion of the pulsar measurements markedly strengthens our analysis, providing leverage on both the inferred diffuse slope $\alpha_1$ and the characteristic density $n_0$, as we shall demonstrate.

We run our analysis on the full data set as well as after removing the outlier to test how impactful it is. Therefore, our study includes four \textbf{D}ata \textbf{S}ets (DS):
\begin{itemize}
  \item DS1: Original Zeeman data \citepalias{Crutcher2010}
  \item DS2: Original Zeeman data with outlier removed
  \item DS3: Expanded Zeeman \citetext{\citetalias{Crutcher2010};  \citealt{Hwang2024}} and pulsar data \citep{Seta2025}
  \item DS4: Expanded Zeeman and pulsar data with outlier removed
\end{itemize} 
For clarity, in this work $n$ denotes the density quantity reported by each probe: tracer-dependent neutral or molecular densities for Zeeman points and LOS-averaged electron number density $n_e$ for pulsar points. Thus, in the combined fits, $n$ is an effective phase-dependent density coordinate rather than a single uniform species definition.

As the compiled values of $n$ depend on observational methodology ($n_e$ for pulsars; $n_{\mathrm{H}}$ or $n_{\mathrm{H_2}}$ proxies for Zeeman tracers), systematics can affect both slope and break estimates. In particular, pulsar points represent LOS-averaged ionized-gas-weighted densities, while Zeeman points are tied to neutral or molecular cloud observations. This disparity can shift the relative normalisation of low- and high-density regimes and broaden or shift the inferred transition density $n_0$. It can also bias the power-law $\alpha_1$ if the effective number density scale for pulsars is systematically offset relative to Zeeman densities. The hyperparameters defined in the next subsection absorb part of this residual dispersion, but do not fully remove structured cross-tracer bias. We therefore treat DS1/DS2 (Zeeman-only) versus DS3/DS4 (Zeeman+pulsar) comparisons, together with outlier tests and Pareto-smoothed importance-sampling leave-one-out cross-validation (PSIS-LOO) model comparisons, as explicit robustness checks on these systematics (Sect.\ \ref{sec:single_slope}). We address the effects of sensitivity limits in detail in Appendix~\ref{ap:sensitivity}.

\subsection{Extended Hierarchical Bayesian Model}
\label{sec:model}
We extend our hierarchical models \citepalias{Whitworth2025} by incorporating several modifications inspired by \citet{Jiang2020}. Because both the Zeeman and pulsar measurements provide LOS-averaged magnetic field strengths that, owing to projection and cancellation effects, represent lower limits to the true field strength, a hierarchical model accounting for these effects and allowing for statistical inference on the upper field envelope is required. 

We adopt a two-part power-law as our favoured model. In order to test the robustness of this choice, in Section \ref{sec:single_slope} we compare our fiducial case against a single-slope model, described by Equation~\ref{power_law}. The two slope model is motivated by the expectation that the $B$--$n$ relation differs between the diffuse and dense ISM regimes traced by the data. These regimes are expected to be governed by different dominant physical processes, including feedback-driven turbulence, shear, thermal effects, and gravitational collapse. It is important to emphasize that the fiducial model permits the slopes in the two regimes to differ, but does not enforce this. If supported by the data, it can recover an effective single power-law solution.

\subsubsection{Model A - Free density correction}

This is our fiducial model. The remaining models are modifications of this baseline and introduce progressively different assumptions.

We retain the same piecewise power-law envelope for the magnetic field introduced in \citetalias{Whitworth2025} and given by Equation \eqref{power_law}, a log-space treatment of priors and parameter space, and Gaussian likelihood in $\log B$. The extended model includes several parameters: $\alpha_1$, $\alpha_2$, $B_0$, $n_0$, and hyperparameters to account for different observational subtleties: the errors in the magnetic field estimates $\sigma_B$, an envelope‑scaling parameter $f_B$, and a shared log‑density intrinsic scatter $R_n$.

We implement a log‑density intrinsic scatter $R_n$ that represents the width of a Gaussian error in log-space on $n$, parameterized by a hyperparameter $f_n$, which represents the typical multiplicative uncertainty in the reported density in linear space. This parameter quantifies the intrinsic scatter in $\log_{10}\, n$. For each source, the model assumes the true density could be higher or lower than the observed value, with a typical factor of uncertainty $f_n$, corresponding to a standard deviation $R_n$ in log space. Thus,
\begin{equation} \label{eq:R}
    R_n = \log_{10}(1+f_n).
\end{equation}

That is, for all sources $i$,
\begin{equation}
    \log_{10} n_{\mathrm{model},i} \sim \mathcal{N} (\log_{10} n_{\mathrm{obs},i}, R_n^2) 
    \label{eq:2}
\end{equation}
where $R_n$ is inferred from the data. This is similar to the $R$ reported in \citet{Jiang2020} and the factor 2 reported in \citetalias{Crutcher2010}.

The envelope-scaling hyperparameter $f_B$ captures the effects of projection and LOS systematics, analogous to the $f$ parameter(s) in \citet{Jiang2020}. However, we assign $f_B$ log-normal rather than uniform priors, which better captures the expected scatter in projection effects. We therefore define the transform in log space as:
\begin{equation}
\label{eq:b_transform}
    \log_{10} B_\mathrm{obs} = \log_{10} B_\mathrm{model} + \log_{10} f_B
\end{equation}
This hyperparameter statistically adjusts the predicted magnetic field strength to account for systematic reductions in observed field strengths relative to the true values, enabling the model to represent the upper envelope of true values statistically, without having to identify it explicitly. 

To capture additional field dispersion, we include a free parameter $\sigma_B$ controlling the intrinsic scatter in $\log_{10} B$ due to line-of-sight averaging through turbulent fields and variation across different regions beyond that captured by the measurement errors.

For each observation $i$, we define $\log_{10} B_{\mathrm{model},i}$ as the deterministic, noise-free magnetic field predicted by the broken power-law envelope at the modelled density $n_{\mathrm{model},i}$. That is, for each $i$, $\log_{10} B_{\mathrm{model},i}$ is given by: 
\begin{multline}
\label{eq:model_A_field}
    \log_{10} B_\mathrm{model,i} = \log_{10} B_0 +\\
    \begin{cases}
    \alpha_1 (\log_{10} n_\mathrm{model,i} - \log_{10} n_0) \mbox{ if } \log_{10} n_\mathrm{model,i} \le \log_{10} n_0, \\
    \alpha_2 (\log_{10} n_\mathrm{model,i} - \log_{10} n_0) \mbox{ if } \log_{10} n_\mathrm{model,i} > \log_{10} n_0.
    \end{cases}
\end{multline}
We interpret $B_{\mathrm{model},i}$ as defining the upper envelope of magnetic-field strengths as a function of density.

We model the total scatter $\sigma_{B,i,\mathrm{total}}$ in a magnetic field observation $B_i$ as a combination of intrinsic and measured components,
\begin{equation}
\sigma_{B,i,\mathrm{total}}^2 = \sigma_B^2 + \sigma_{B,i}^2,
\end{equation}
where $\sigma_{B,i}$ is the per-source log-space observational uncertainty of the measured field, and $\sigma_B$ is the intrinsic scatter inferred as a hyperparameter. 

The log-likelihood for the observed log-field values $\log_{10} B_{obs,i}$ and set of hyperparameters $\theta$ is:
\begin{multline}
\label{eq:likelihood}
    \ln p({B_{\mathrm{obs,i}}} \mid \theta) = \\
    -\frac{1}{2} \sum_{i=1}^{n} \Biggl[
\left( \frac{\log_{10} B_{\mathrm{obs,i}} - (\log_{10} B_\mathrm{model,i} + \log_{10} f_B)}{\sigma_{B,i,\mathrm{total}}} \right)^2 \\
+ \ln \sigma_{B,i,\mathrm{total}}^2 \Biggr] + \mathrm{const},
\end{multline}
where $(B_{\mathrm{model},i}(\theta))$ is the deterministic envelope prediction from Equation~\eqref{eq:model_A_field}. This log-likelihood function, allows the observed values to lie below this curve due to the envelope scaling factor $f_B$ along with additional intrinsic scatter.

The log-likelihood assumes that the observed magnetic field $B_{obs,i}$ is drawn from a normal distribution centred on the predicted value ($\log_{10} B_{\mathrm{model},i}(\theta)$), with total variance $\sigma_{B,i}^2 $ (intrinsic plus measurement uncertainty). Thus, the agreement between each measured $\log_{10}B_i$ and predicted $\log_{10} B_{\mathrm{model},i}$ from the modelled is rigorously assessed.

We implement the models in a fully vectorized log-likelihood evaluation for computational efficiency. The full set of model hyperparameters and priors for each model is given in Table~\ref{table:all_models}. We note that some hyperparameters and priors are not in log space: specifically, $\alpha_1$, $\alpha_2$ and $\sigma_B$ are defined in linear space. This choice reflects our expectation that these parameters do not span many orders of magnitude, and therefore, a linear-space treatment is more appropriate. We assign log space Gaussian priors to the scale parameters, $n_0$, $B_0$, $f_B$, as well as the density intrinsic scatter parameter $f_n$.

We examine the effects of sensitivity limits in detail in Appendix \ref{ap:sensitivity}, and show that accounting for these limits can appreciably affect the inferred $B$--$n$ relationship. However, restricting the sample based on sensitivity reduces the number of usable data points, which introduces bimodality in the posterior distributions and makes it more difficult for the Bayesian analysis to constrain the relationship. Fully incorporating sensitivity effects would require a more detailed hierarchical framework \citep{Loredo2004,Kelly2007}, which is far beyond the scope of this work. Therefore, we include all reported data as given in our analysis.

\subsubsection{Model B - Fixed density correction}

Our second model is identical to the fiducial model except that instead of inferring the scatter width in log-density ($R_n$) from the data, we fix $R_n$ to a set of prescribed values. Specifically, we adopt:
\begin{equation}
\log_{10} n_{\mathrm{model},i} \sim \mathcal{N}(log_{10}n_{{obs},i}, R_n),
\end{equation} 
where $R_n$ is fixed to 0.30, 0.70, or 0.95, corresponding to multiplicative factors $f_n$ = 2, 5, 9. These choices allow direct comparison with the uncertainty factors assumed by \citetalias{Crutcher2010}, \citetalias{Whitworth2025}, and \citet{Jiang2020}. In this model, the typical log-density error is not marginalized over, but kept constant for all sources.

\subsubsection{Model C - Two-envelope model with flexible density correction}

Having tested the new approach with Models~A and~B, we next improve upon them by taking Model~A and generalising the single magnetic field envelope-scaling parameter $f_B$ into two parameters, $f_{B,1}$ and $f_{B,2}$, applied respectively below and above $n_0$. This is motivated by the expectation that the magnetic field in diffuse gas behaves differently from that in dense gas \citep[][\citetalias{Whitworth2025}]{Jiang2020} due to different density and turbulent velocity properties \citep{Seta2025}, and by the possibility that projection effects, field tangling, and/or beam averaging differ systematically between these regimes due to different size and morphology of structures. Both $f_{B,1}$ and $f_{B,2}$ are treated as global hyperparameters and inferred simultaneously from the data. 

To obtain a well-behaved Bayesian model and ensure convergence, we slightly modify our priors and bounds relative to Models~A and~B. The additional levels of complexity in this model increase parameter degeneracies, and overly wide priors can therefore prevent the sampling routine from converging. This results in having to adopt narrower priors informed by the posterior constraints from Models~A and~B.

\subsubsection{Model D - Two-envelope model with reported pulsar density errors}

Model~D uses the same hyperparameter set as Model~C but modifies the treatment of density errors for the pulsar data by incorporating the reported density errors. Unlike the Zeeman measurements, the pulsar density estimates can include error estimates via the propagation of dispersion measure uncertainties as well as distance uncertainties when available \citep[out of 212 pulsars, distance uncertainties are available for 126 pulsars only; see][]{Manchester2005}. By including these reported errors, the Bayesian framework can more realistically account for measurement limitations in pulsar-derived densities. However, because these errors are relatively small compared to the systematic uncertainties, we retain the log-density intrinsic scatter.

For each data point $i$ we modify Equation \ref{eq:2}:
\begin{equation}
\log_{10} n_{\mathrm{model},i} \sim \mathcal{N}(\log_{10} n_{obs,i}, n_{\mathrm{eff},i}^2),
\end{equation}
where the effective error, $n_{\mathrm{eff},i}$ is given by the quadrature sum of the log-space Gaussian intrinsic scatter $R_n$ and the reported error:
\begin{equation}
\label{eq:pulsar_err}
n_{\mathrm{eff},i} = \sqrt{R_n^2 + (\log_{10} \sigma_{n,i})^2 },
\end{equation}
and $\sigma_{n,i}$ is the reported fractional error on $n$ for each pulsar measurement. For the pulsar data, we use the reported values of $n_{\mathrm{err},i}$; for the Zeeman data, we set $n_{\mathrm{err},i}=0$, so that the total uncertainty there is then governed entirely by $f_n$. As in the previous models, we report $R_n$ as given by Equation~\eqref{eq:R}.

In this model, $f_n$ contributes as an effective systematic shift that combines with the reported measurement errors in quadrature, providing a minimum correction when no errors are reported (as in Models~A and~C). Allowing for two envelope parameters and the additional error propagation increases the flexibility of the model but also amplifies parameter degeneracies. We therefore adopt the same limits and priors as in Model~C. As this only affects the pulsar data sets, we only run the model on DS3 and DS4.

\begin{table*}
\begin{center}
\caption{Model hyperparameters and priors}
\begin{tabular}{llccc}
\hline \\
 & \textbf{Parameter} & \multicolumn{3}{c}{\textbf{Model}} \\
 & & A & B & C/D \\
\hline \hline
\\
$B_0$ [$\log_{10} B_0$] & Envelope normalization ($\mu$G) & 
  $\log_{10} B_0 \sim \mathcal{N}(0,2)$ & as A & as A \\ 
 & & $>-10-Y_\mathrm{min}$ & & \\ 
\\
$n_0$ [$\log_{10} n_0$] & Break density (cm$^{-3}$) & 
  $\log_{10} n_0 \sim \mathcal{N}(0,2)$ & as A & as A \\ 
 & & $>-10-X_\mathrm{min}$ & & \\ 
\\
$\alpha_1$ & Low-$n$ slope & $\mathcal{N}(0.5,0.5)$ & as A & $\mathcal{N}(0.2,0.2)$ \\ 
 & & $>0$ & & $[0,1]$ \\ 
\\
$\alpha_2$ & High-$n$ slope & $\mathcal{N}(0.5,0.5)$ & as A & $\mathcal{N}(0.6,0.2)$ \\ 
 & & $>0$ & & $[0,1]$ \\ 
\\
$f_B$ [$\log_{10} f_B$] & Envelope scaling & $\log_{10} f_B \sim \mathcal{N}(0,1.0)$ & as A & -- \\ 
 & & $[10^{-4},1]$ & & -- \\ 
\\
$f_{B,1}$ [$\log_{10} f_{B,1}$] & Low-$n$ envelope scaling & -- & -- & $\log_{10} f_{B,1} \sim \mathcal{N}(0,0.1)$ \\ 
 & & -- & -- & $[10^{-4},1]$ \\ 
\\
$f_{B,2}$ [$\log_{10} f_{B,2}$] & High-$n$ envelope scaling & -- & -- & $\log_{10} f_{B,2} \sim \mathcal{N}(0,0.1)$ \\ 
 & & -- & -- & $[10^{-4},1]$ \\ 
\\
$\sigma_B$ & $B$ intrinsic scatter & $\mathcal{N}(0,1.0)$ & as A & $\mathcal{N}(0.1,0.1)$ \\ 
 & & $>0$ & & $[0,10]$ \\ 
\\
$f_n$ [$\log_{10} f_n$] & intrinsic scatter 
& $\log_{10} f_n \sim \mathcal{N}(0,1.0)$ & $(f_n + 1)$ fixed as: 2, 5, 9 & as A \\
 & reported as $R_n = \log_{10}(1+f_n)$ & $[10^{-2},10^3]$ & & \\
\\
\hline \hline
\label{table:all_models}
\end{tabular}
\end{center}
\par\noindent
\textit{Notes.} The top line for each parameter shows the prior choice whilst the bottom shows the HMC bounds. Model~B fixes $R_n$ to literature values. C/D use dual $f_{B,1}$, $f_{B,2}$.
\end{table*}

\subsection{Posterior Sampling and Chains}
We use a Hamiltonian Monte Carlo (HMC) approach to sample posteriors. As an improvement of the traditional Metropolis-Hastings Markov Chain Monte Carlo \citep[MCMC][]{Metropolis1953} approach, the HMC algorithm uses gradient information to take informed steps, reducing random-walk behaviour and therefore requiring fewer samples to reach convergence \citep{Neal2011, Betancourt2017}. We use the HMC algorithm due to its ability to explore efficiently high-dimensional parameter spaces.

Posterior samples are generated using HMC \citep{Neal2011}, as implemented in \texttt{Stan}, a statistical modelling language\footnote{https://mc-stan.org/} accessed from a Python wrapper, specifically version 1.2.5 of the package \texttt{CmdStanPy}\footnote{https://mc-stan.org/cmdstanpy/}. In particular, we use the No-U-Turn Sampler with parameters \texttt{adapt\_delta} $=0.9$ and \texttt{max\_treedepth} $=12$ tuned to optimize acceptance rate and mixing.

In all models, we use four independent chains, which is a commonly adopted standard for MCMC sampling. It allows for robust convergence diagnostics while balancing computational cost and efficiency. Each chain is initialised from pseudo-random initial values to robustly explore the parameter space. Each chain consists of $5\,000$ warm-up iterations followed by $5\,000$ sampling iterations.

Chain convergence is assessed using the potential scale reduction factor $\hat{R}$ \citep{GelmanRubin1992} and the effective sample sizes for bulk and tail posterior distributions, both computed automatically by the HMC implementation. Note that $\hat{R}$ is distinct from the intrinsic scatter parameter $R_n$; it diagnoses and convergence by comparing between and within-chain variability across independent MCMC chains. For an $\hat{R}$ close to unity, all chains are considered to have mixed well and converged to the stationary posterior distribution. Whilst convergence is typically achieved quickly, our choice of number of sampling iterations per chain ensures a large sample size across all parameters, minimises variability due to random initialisations, and ensures thorough mixing. Posterior samples are then used to construct credible intervals, medians, maximum-a-posteriori (MAP) estimates, and corner plots for each model and dataset.

\section{Results} 
\label{sec:results}

In this section we present the results of our extended hierarchical Bayesian approach that includes the original and extended data sets. We focus on the MAP results as these represent the parameter values at the peak of the posterior distribution (i.e., something like a best fit in a Bayesian sense).

\subsection{Model A - Free density correction}

\begin{figure*}
\centering
\subfloat[DS1: Zeeman data only.]{\includegraphics[scale=0.35]{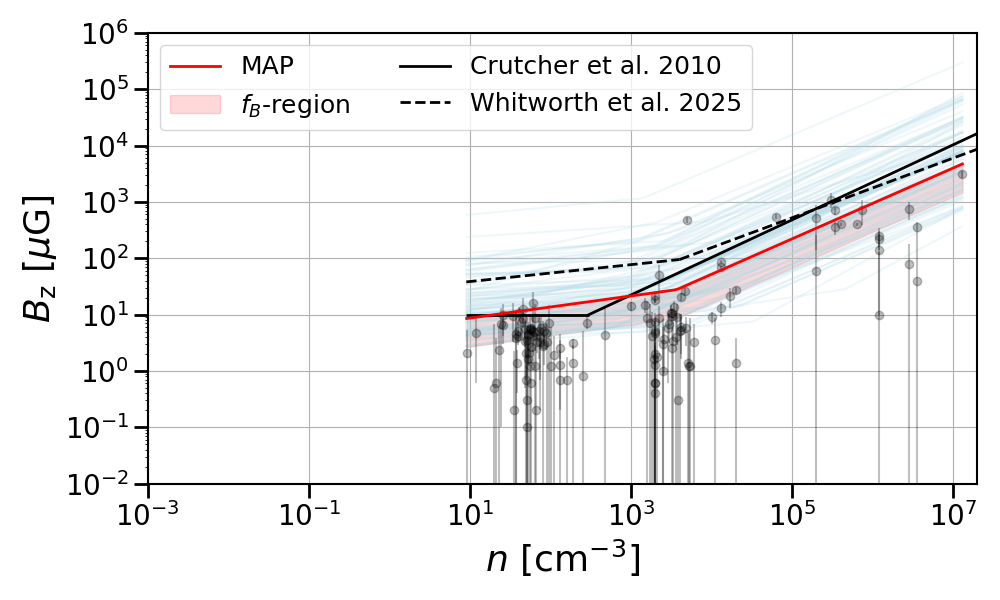}}
\subfloat[DS2: Zeeman data with outlier removed.]{\includegraphics[scale=0.35]{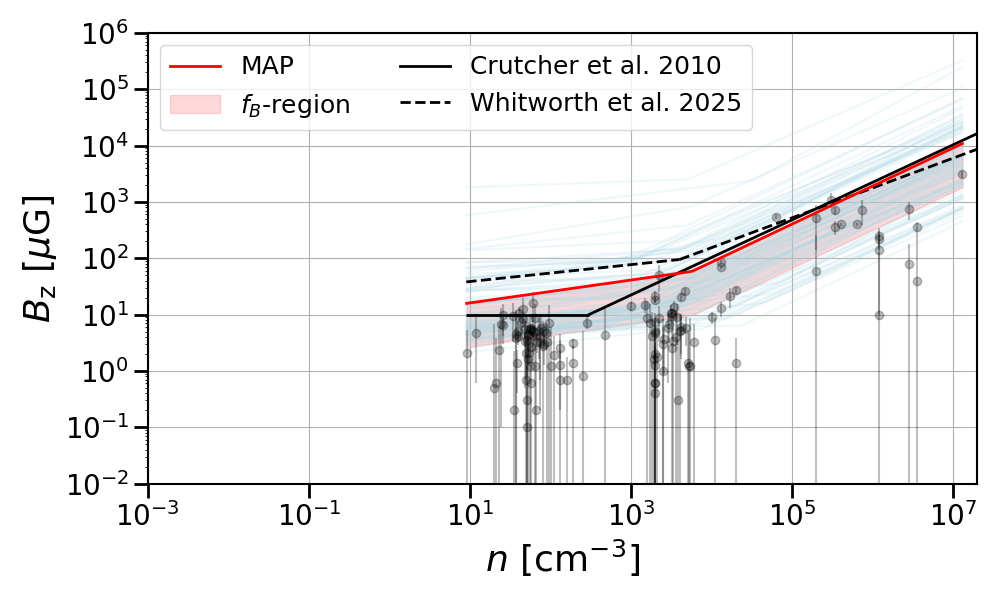}}\\
\subfloat[DS3: Zeeman plus pulsar data.]{\includegraphics[scale=0.35]{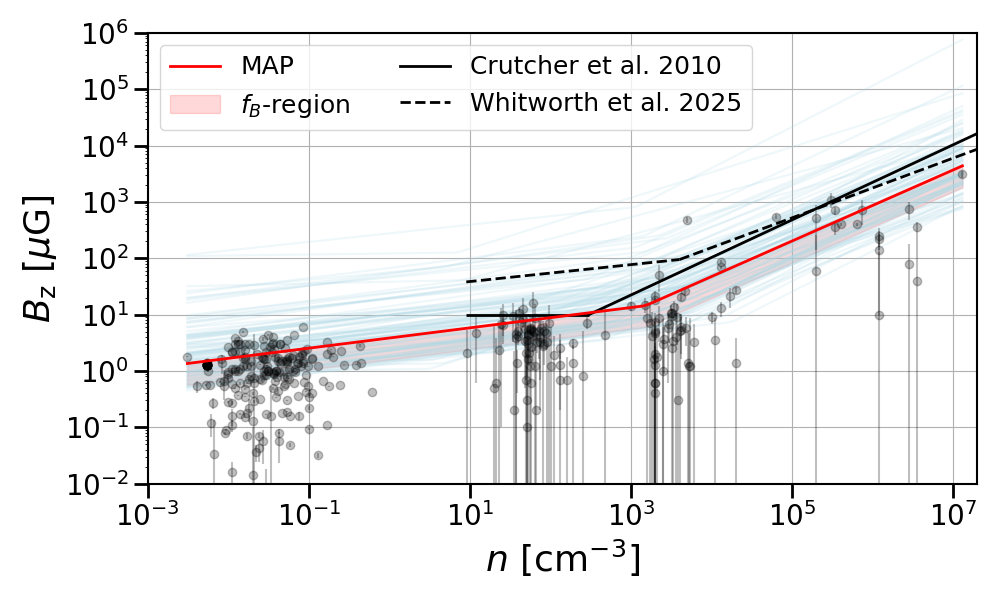}}
\subfloat[DS4: Zeeman plus pulsar, outlier removed.]{\includegraphics[scale=0.35]{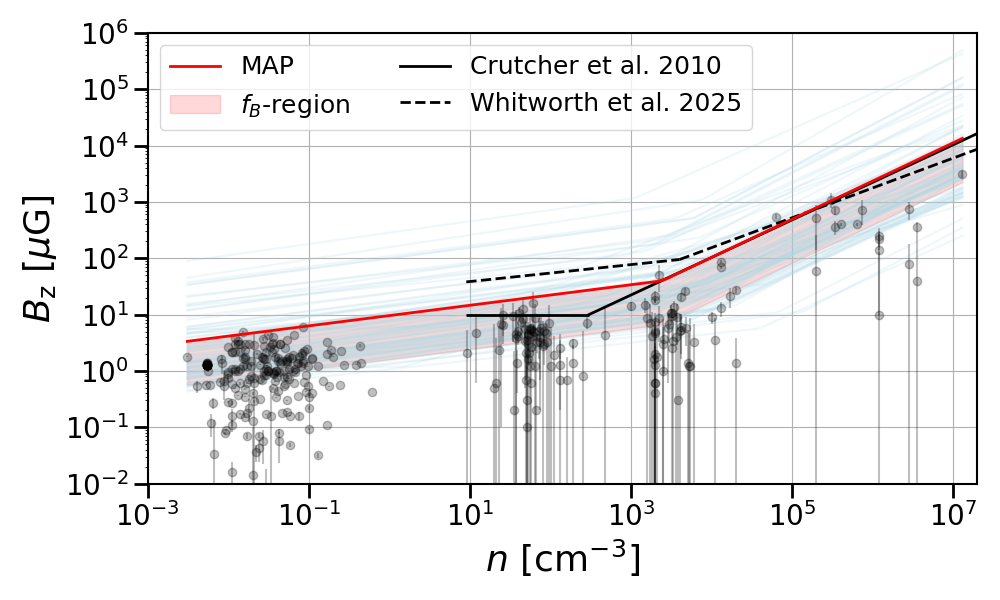}}\\
\caption{The inferred upper limit of magnetic field strength as a function of number density, based on hierarchical Bayesian analysis of Model~A for each dataset (DS1–DS6: see subfigure captions). The {\em solid black line} in all plots is the upper envelope proposed by C10. The {\em red line} shows the MAP relationship between $B_z$ and $n$ based on our analysis, and the {\em blue lines} are 100 random posterior draws. The scaling envelope of $f_B$, shown as the {\em pink band}, represents how far $B$ could plausibly lie below the MAP. The dashed line shows the result published in \citetalias{Whitworth2025}, which had an incorrectly high value of $B_0$, as is further discussed in Appendix~\ref{Append:code}. Error bars for $n$ are omitted for readability.}
\label{fig:model_A_grid}
\end{figure*}

Figure \ref{fig:model_A_grid} and Table \ref{table:model_a_results} show the results for the inferred MAP results for each dataset, with the $68\%$ highest posterior density (HPD).

Across all data sets, $\alpha_1$ is well constrained, varying by $\leq 10\%$ between $0.18^{+0.02}_{-0.03}$ to $0.20^{+0.03}_{-0.20}$ The value for $\alpha_2$ increases from $0.63$ to $0.68$ when the outlier is removed, consistent with \citet{Jiang2020}, although this is only just within the $68\%$ HPD.

The break density, $n_0$, varies between datasets, from $1.51^{+0.71}_{-1.51} \times 10^3 \rm\,cm^{-3}$ to $5.71^{+0.89}_{-5.30} \times 10^3 \rm\,cm^{-3}$. The extended datasets DS3 and DS4 favour lower values of $n_0$ than the Zeeman-only datasets, though the uncertainties remain large and asymmetric. The inferred values are larger than reported in \citetalias{Crutcher2010} of $300\rm\,cm^{-3}$, but are comparable with \citetalias{Whitworth2025}.

The inferred values of $B_0$ are larger when the pulsar data is included, ranging from $14.41_{-13.39}^{+11.19}\,\mu$G to $58.53_{-54.10}^{+18.13}\,\mu$G. In both the Zeeman-only and the extended Zeeman plus pulsar data, $B_0$ increases when the outlier is removed.\footnote{We do not compare to the result in \citetalias{Whitworth2025} here, because whilst developing the new code for this paper, an error in the original code was found that lead to an overprediction of $B_0$. See Appendix \ref{Append:code} for a discussion on this point.}

The estimated log-density intrinsic scatter is consistent at around 1 for the Zeeman-only datasets DS1 and DS2 and around 0.45 for DS3 and DS4. Although uncertainties on $R_n$ are large, the central values are in line with those from earlier studies with the Zeeman only data sets matching \citetext{\citealt{Jiang2020}, and the Zeeman plus plusar data sets closer to \citetalias{Whitworth2025}}. This suggests that the size of the intrinsic scatter is dependent on how the data are handled and how the model is implemented.

The intrinsic scatter in magnetic field strength, $\sigma_B$, is small and consistent in the extended datasets DS3 and DS4. The Zeeman-only datasets DS1 and DS2 show a slight decrease when the outlier is removed. The scaling envelope $f_B$, shown in Figure \ref{fig:model_A_grid}, represents the range over which $B$ could plausibly lie, with the MAP being the upper limit. Because $f_B$ acts additively in log space, smaller values correspond to a larger downward offset, thus a larger shaded region on the plot. Table \ref{table:model_a_results} shows that the Zeeman-only results DS1 and DS2 yield a smaller $f_B$ compared to the combined datasets, while removing the outlier slightly reduces $f_B$. This indicates a larger inferred suppression of the observed field strength in the Zeeman-only samples.

\begin{table*}
    \begin{center}
    \caption{MAP values from model~A.}
    \begin{tabular}{lccccccc}  
    \hline
    \\
        Data & $\alpha_1$ & $\alpha_2$ & $n_0$ & $B_0(\mu{\rm G})$ & $f_B$ & $\sigma_B$ & $R_n$\\
        Set & && $(10^3{\rm cm}^{-3})$ &&&&
        \\
        \\
        \hline \hline 
        \\
        DS1 &  $0.20_{-0.20}^{+0.03}$ & $0.63_{-0.06}^{+0.04}$ & $3.66_{-3.36}^{+0.13}$ & $27.69_{-23.47}^{+11.12}$ & $0.31_{-0.31}^{+0.19}$ & $0.31_{-0.02}^{+0.04}$ & $0.97_{-0.96}^{+0.46}$ \\
        \\
        DS2 &  $0.20_{-0.20}^{+0.03}$ & $0.68_{-0.08}^{+0.02}$ & $5.71_{-5.30}^{+0.89}$ & $58.53_{-54.10}^{+18.13}$ & $0.17_{-0.16}^{+0.35}$ & $0.26_{-0.02}^{+0.03}$ & $1.11_{-1.10}^{+0.32}$  \\ 
        \\
        DS3 &  $0.18_{-0.03}^{+0.02}$ & $0.63_{-0.09}^{+0.06}$ & $1.51_{-1.51}^{+0.71}$ & $14.41_{-13.39}^{+11.19}$ & $0.42_{-0.41}^{+0.14}$ & $0.44_{-0.02}^{+0.02}$ & $0.45_{-0.44}^{+0.41}$ \\   
        \\
        DS4 &  $0.18_{-0.02}^{+0.02}$ & $0.68_{-0.11}^{+0.04}$ & $2.45_{-2.45}^{+1.35}$ & $39.90_{-36.60}^{+7.80}$ & $0.17_{-0.16}^{+0.39}$ & $0.42_{-0.02}^{+0.02}$ & $0.46_{-0.45}^{+0.36}$  \\
        \\
        \hline \hline
        \label{table:model_a_results}
   \end{tabular}
   \end{center}
   \par\noindent\textit{Notes.} 
     $R_n$ is the inferred log-density intrinsic scatter for the observed number density: it is inferred in Models A, C, and D. $f_B$ is the multiplicative factor scaling the envelope. $\sigma_B$ is the fractional uncertainty that is applied to the errors on $B$. DS1 is the original Zeeman data, DS2 is the Zeeman data with its outlier removed, DS3 is the Extended Zeeman data with pulsar data added, and DS4 is the Zeeman plus pulsar data with both outliers removed. The errors are the bounds of the 68th percentile confidence interval.   
\end{table*}

\subsection{Model B - Fixed density correction}

\begin{figure*}
\centering
\subfloat[DS1: Zeeman data only.]{\includegraphics[scale=0.35]{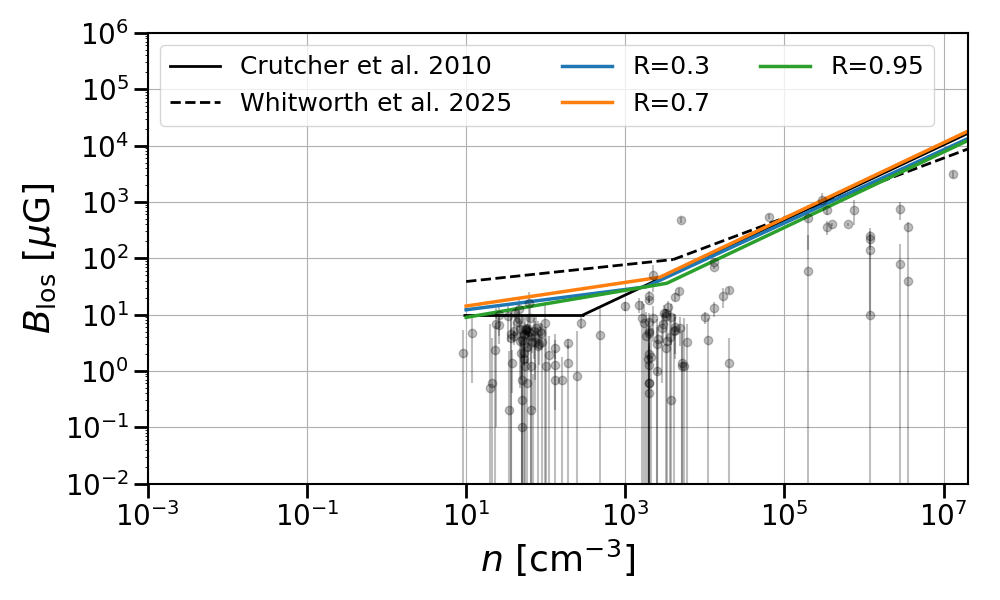}}
\subfloat[DS2: Zeeman data with outlier removed.]{\includegraphics[scale=0.35]{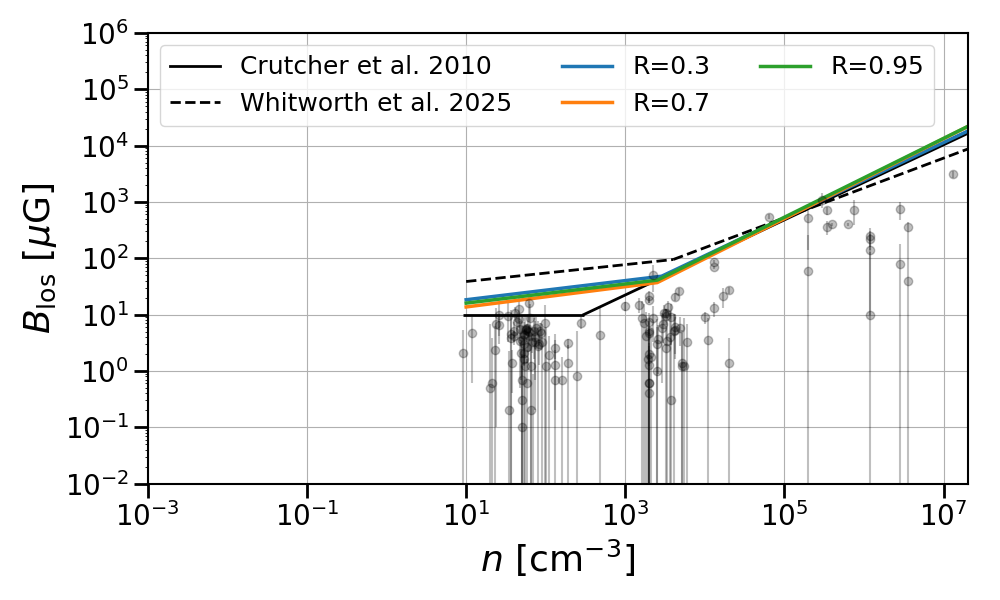}}\\
\subfloat[DS3: Zeeman plus pulsar data.]{\includegraphics[scale=0.35]{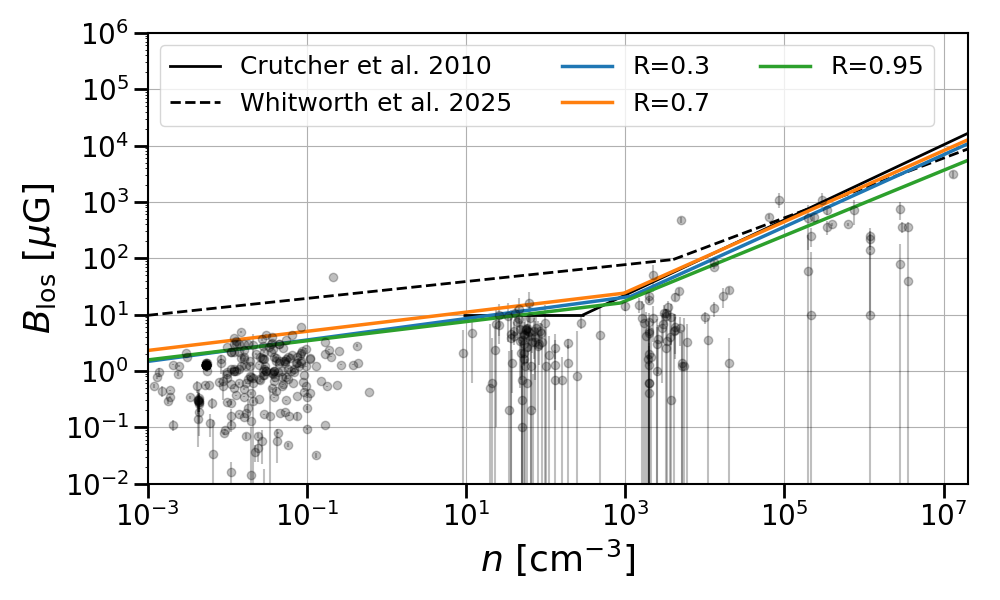}}
\subfloat[DS4: Zeeman plus pulsar, outlier removed.]{\includegraphics[scale=0.35]{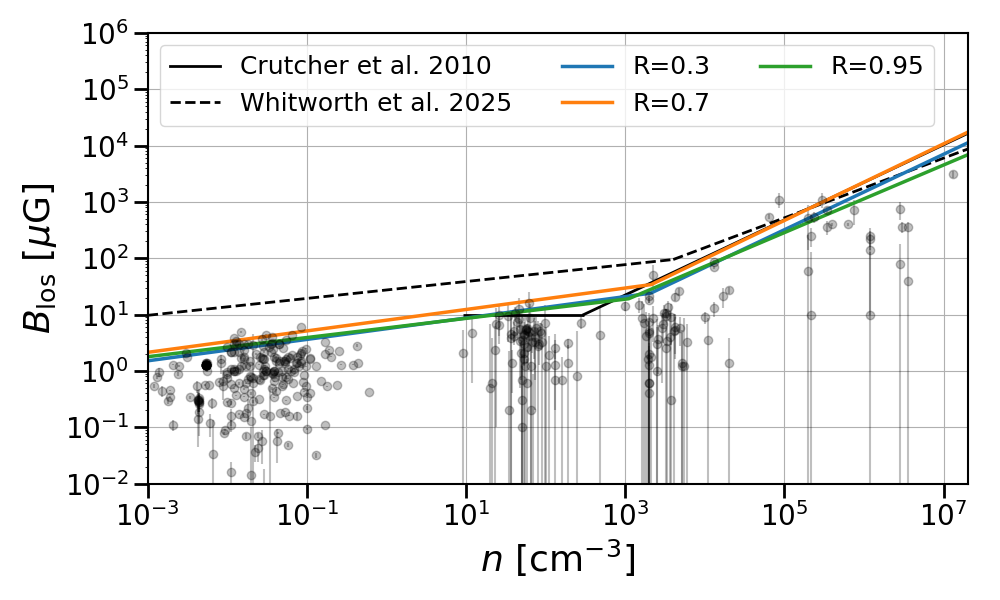}}\\
\caption{The inferred relationship of magnetic field strength as a function of number density, based on hierarchical Bayesian analysis for each dataset for Model B, with fixed linear density error parameter $(f_n + 1)$. Here we plot the results for the different values of $(f_n + 1)$ on the same plot for each dataset for ease of comparison between the error treatments. Otherwise the same as Fig.~\ref{fig:model_A_grid}.}
\label{fig:model_B_grid}
\end{figure*}

Table~\ref{table:model_b_results} presents the MAP results for Model B, while Figure~\ref{fig:model_B_grid} shows the MAP fits for each data set at each fixed value of $R_n$ (for readability, we do not plot the 100 sample lines or $f_B$ region).

Across all datasets and values of $R_n$, $\alpha_1$ is well constrained, with typical values of $\sim 0.19$. $\alpha_2$ also remains consistent, but shows some variation due to the outlier. In contrast $n_0$ shows a strong dependency on $R_n$. As $R_n$ increases, both $n_0$ and its uncertainties increase. For all values of $R_n$, the extended datasets DS3 and DS4 favour lower $n_0$, showing the influence of the low-density pulsar data, even at high fractional uncertainty.

The values for $B_0$, $f_B$, and $\sigma_B$ show no trend with $R_n$, inclusion of the outlier, or the pulsar data. They fluctuate across the different fits, suggesting that Model~B is less sensitive to these parameters or that the heterogeneous data do not support a single simple trend for them.

\begin{table*}
    \begin{center}
    \caption{The MAP results for Model~B}
    \begin{tabular}{lccccccccc}  
    \hline
    \\
        Data&  $\alpha_1$ & $\alpha_2$ & $n_0$ & $B_0 (\mu{\rm G})$ & $f_B$ & $\sigma_B$ & $R_n$ & $(f_n + 1)$ \\
        Set& &&$(10^3{\rm cm}^{-3})$ &&&&&&
        \\
        \\
        \hline \hline 
        \\
        DS1 & $0.18_{-0.18}^{+0.02}$ & $0.65_{-0.07}^{+0.03}$ & $1.79_{-1.52}^{+0.28}$ & $30.70_{-26.73}^{+6.46}$ & $0.27_{-0.27}^{+0.25}$ & $0.16_{-0.13}^{+0.01}$ & 0.30 & 2 \\
        \\
        DS2 & $0.17_{-0.14}^{+0.03}$ & $0.67_{-0.07}^{+0.03}$ & $2.86_{-2.48}^{+0.35}$ & $47.85_{-43.34}^{+5.62}$ & $0.19_{-0.19}^{+0.32}$ & $0.08_{-0.08}^{+0.00}$ & 0.30 & 2 \\
        \\
        DS3 & $0.19_{-0.03}^{+0.02}$ & $0.64_{-0.10}^{+0.06}$ & $1.17_{-1.17}^{+0.12}$ & $20.75_{-19.89}^{+3.81}$ & $0.31_{-0.30}^{+0.25}$ & $0.31_{-0.02}^{+0.03}$ & 0.30 & 2 \\
        \\       
        DS4 & $0.19_{-0.03}^{+0.01}$ & $0.67_{-0.09}^{+0.05}$ & $2.08_{-1.90}^{+0.32}$ & $24.30_{-20.54}^{+6.49}$ & $0.31_{-0.29}^{+0.27}$ & $0.30_{-0.03}^{+0.03}$ & 0.30 & 2 \\
        \\
        \hline \hline
        \\
        DS1 & $0.21_{-0.21}^{+0.02}$ & $0.67_{-0.19}^{+0.07}$ & $2.71_{-2.71}^{+0.52}$ & $46.25_{-46.24}^{+14.21}$ & $0.19_{-0.18}^{+0.33}$ & $0.07_{-0.07}^{+0.02}$ & 0.70 & 5 \\
        \\
        DS2 & $0.18_{-0.17}^{+0.03}$ & $0.71_{-0.18}^{+0.06}$ & $2.52_{-2.52}^{+0.74}$ & $26.87_{-26.87}^{+8.74}$ & $0.26_{-0.26}^{+0.25}$ & $0.09_{-0.09}^{+0.01}$ & 0.70 & 5 \\ 
        \\
        DS3 & $0.17_{-0.06}^{+0.04}$ & $0.63_{-0.26}^{+0.01}$ & $0.96_{-0.96}^{+0.07}$ & $23.68_{-22.88}^{+5.64}$ & $0.22_{-0.21}^{+0.33}$ & $0.04_{-0.04}^{+0.01}$ & 0.70 & 5 \\
        \\
        DS4 & $0.19_{-0.05}^{+0.02}$ & $0.68_{-0.23}^{+0.06}$ & $2.10_{-2.10}^{+0.49}$ & $33.50_{-32.64}^{+11.49}$ & $0.21_{-0.20}^{+0.34}$ & $0.05_{-0.05}^{+0.01}$& 0.70 & 5 \\
        \\
        \hline \hline
        \\
        DS1 & $0.24_{-0.24}^{+0.19}$ & $0.67_{-0.33}^{+0.06}$ & $3.32_{-3.32}^{+2.21}$ & $36.32_{-36.32}^{+16.34}$ & $0.25_{-0.24}^{+0.24}$ & $0.11_{-0.11}^{+0.00}$ & 0.95 & 9 \\
        \\
        DS2 & $0.17_{-0.17}^{+0.18}$ & $0.70_{-0.34}^{+0.03}$ & $2.54_{-2.54}^{+0.66}$ & $41.37_{-41.37}^{+18.85}$ & $0.16_{-0.16}^{+0.34}$ & $0.10_{-0.10}^{+0.00}$ & 0.95 & 9 \\
        \\
        DS3 & $0.17_{-0.15}^{+0.02}$ & $0.58_{-0.35}^{+0.09}$ & $0.86_{-0.86}^{+0.77}$ & $15.57_{-15.56}^{+5.57}$ & $0.33_{-0.32}^{+0.19}$ & $0.07_{-0.07}^{+0.00}$ & 0.95 & 9 \\ 
        \\       
        DS4 & $0.17_{-0.14}^{+0.03}$ & $0.60_{-0.37}^{+0.07}$ & $1.09_{-1.09}^{+0.82}$ & $19.02_{-19.00}^{+7.33}$ & $0.27_{-0.27}^{+0.25}$ & $0.06_{-0.06}^{+0.00}$ & 0.95 & 9 \\
        \\
        \hline \hline
    \label{table:model_b_results}
   \end{tabular}  
   \end{center}
   \par\noindent\textit{Notes.} 
     The same headings and labels as Table \ref{table:model_a_results}, except for the last column, which gives the value for the linear intrinsic scatter corresponding to $R_n$ in each case. 
\end{table*}

\subsection{Model C - Two-envelope model with flexible density correction}

\begin{figure*}
\centering
\subfloat[DS1: Zeeman data only.]{\includegraphics[scale=0.34]{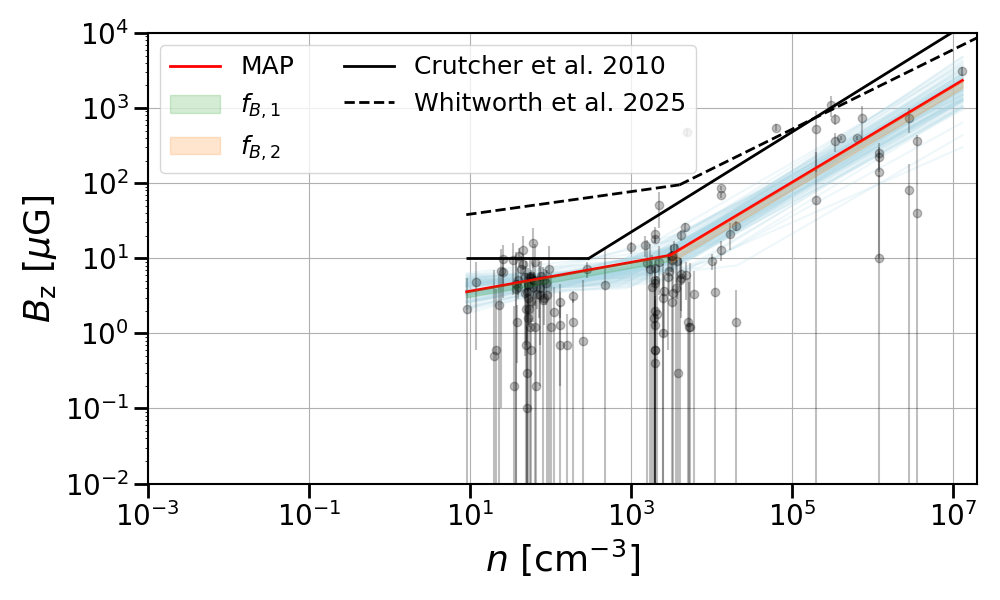}}
\subfloat[DS2: Zeeman data with outlier removed.]{\includegraphics[scale=0.34]{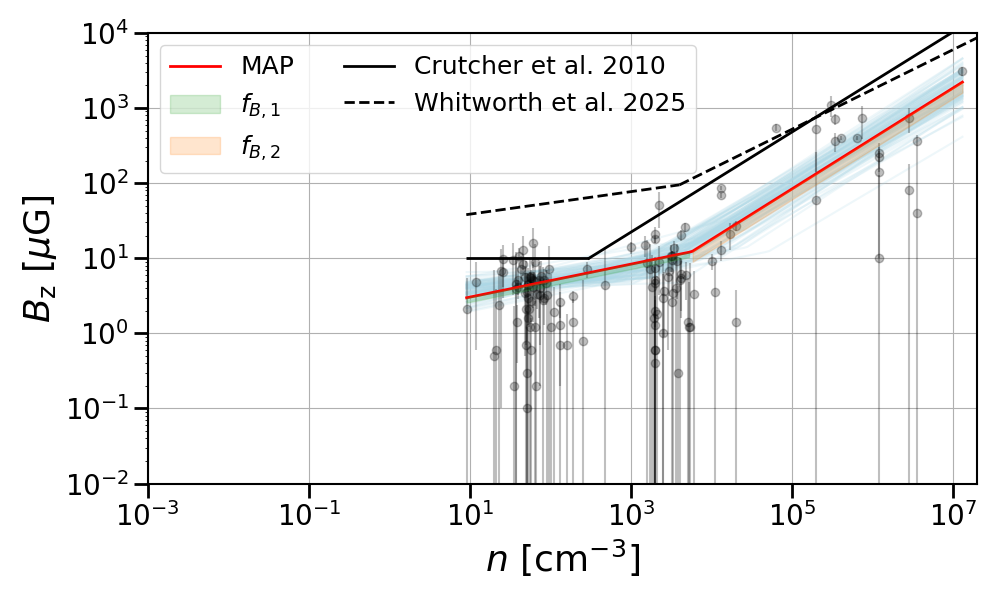}}\\
\subfloat[DS3: Zeeman plus pulsar data.]{\includegraphics[scale=0.34]{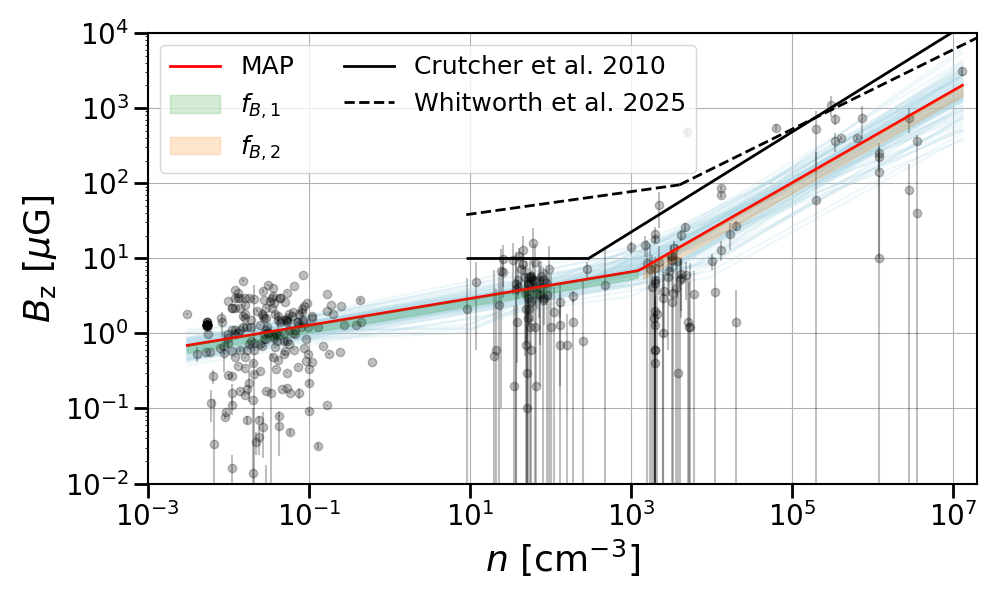}}
\subfloat[DS4: Zeeman plus pulsar, outlier removed.]{\includegraphics[scale=0.34]{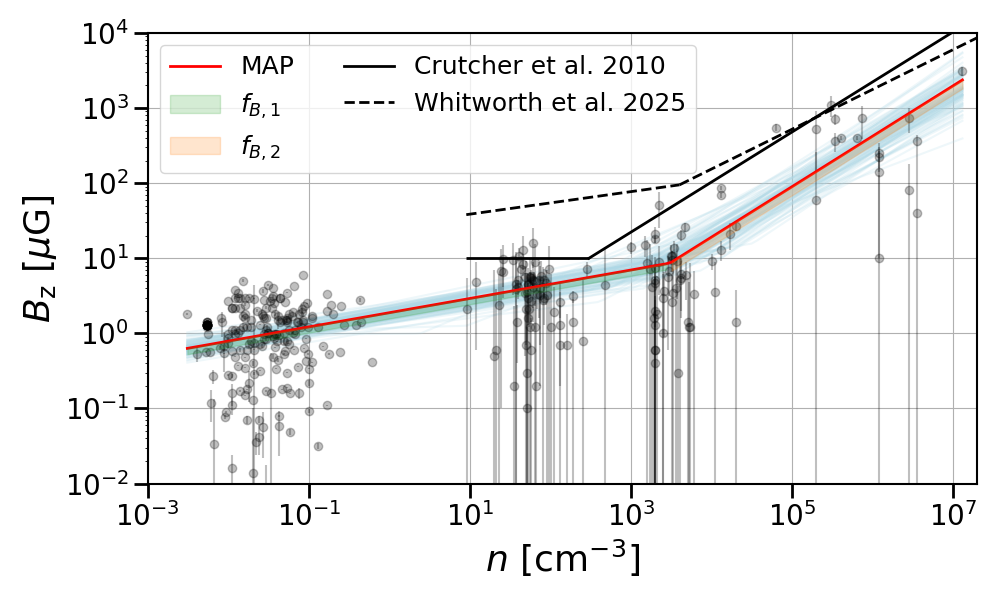}}\\
\caption{MAP results for model C with separate values of $f_B$ above and below $n_0$. The error ranges for $f_{B,1}$ (green) and $f_{B,2}$ (orange) are shown. Otherwise the same as Fig.~\ref{fig:model_A_grid}.}
\label{fig:model_C_grid}
\end{figure*}

Table~\ref{table:model_c_results} presents the MAP results for Model~C, while Figure~\ref{fig:model_C_grid} shows the MAP fits for each dataset, with the two envelope scaling factors $f_{B,1}$ and $f_{B,2}$ shown.

The value of $\alpha_1$ has moderate variation, from $0.18_{-0.02}^{+0.02}$ in DS3 to $0.22_{-0.19}^{+0.01}$ in DS2, with DS1 and DS4 both $\sim 0.19$. For the Zeeman only data sets DS1 and DS2, $\alpha_1$ had asymmetric credible intervals, with the lower interval the larger. Once pulsar data is included in DS3 and DS4, the credible intervals are much more constrained and symmetric. The distributions of $\alpha_2$ are dependent on dataset with DS1, DS2, and DS4 showing asymmetric distributions, whilst DS3 is symmetric but still moderately broad. The value of $\alpha_2$ does, however, lie in a consistent range of $\sim0.61$--$0.67$.

Break density $n_0$ follows a trend similar to Models~A and~B, increasing when the outlier is removed \and higher than the extended datasets. In the extended datasets, DS3 and DS4, $n_0$ also increases with the exclusion of the outlier, but remains consistently lower than the Zeeman-only datasets. dependency on the outlier. 
The envelope normalization, $B_0$, is broadly consistent between datasets and with the value originally reported in \citetalias{Crutcher2010}, as well as being consistent with Appendix \ref{Append:code}, but lower than \citetalias{Whitworth2025}.

The two envelope scaling factors $f_{B,1}$ and $f_{B,2}$, are comparable across datasets with only slight variations, suggesting the two-envelope approach is able to accommodate the observed range of magnetic field strengths across both low- and high-density regimes. The estimated $R_n$ is highly dependent upon the dataset, increasing in the primary datasets when the outlier is removed, yet decreasing in the extended datasets with its removal. This indicates that the required density correction is sensitive to dataset composition and outlier treatment. The intrinsic scatter $\sigma_B$ for DS1, DS2 and DS3 is consistent with Model~A. However, DS4 shows a marked increase, suggesting that in this case a larger fraction of the field variation is being absorbed as intrinsic scatter.

\begin{table*}
    \begin{center}
    \caption{MAP values from model~C.}
    \begin{tabular}{lcccccccc}  
    \hline
    \\
        Data &  $\alpha_1$ & $\alpha_2$ & $n_0$ & $B_0(\mu{\rm G})$ & $f_{B,1}$ & $f_{B,2}$ & $\sigma_B$ & $R_n$ \\
        Set & && $(10^3{\rm cm}^{-3})$ &&&&&
        \\
        \\
        \hline \hline 
        \\
        DS1 & $0.19_{-0.19}^{+0.03}$ & $0.64_{-0.07}^{+0.02}$ & $3.05_{-2.84}^{+0.14}$ & $10.95_{-6.70}^{+1.64}$ & $0.84_{-0.03}^{+0.16}$ & $0.79_{-0.01}^{+0.21}$ & $0.30_{-0.02}^{+0.03}$& $0.67_{-0.63}^{+0.76}$ \\
        \\
        DS2 & $0.22_{-0.19}^{+0.01}$ & $0.67_{-0.07}^{+0.02}$ & $5.69_{-5.29}^{+0.68}$ & $12.24_{-7.76}^{+1.40}$ & $0.86_{-0.05}^{+0.14}$ & $0.72_{-0.06}^{+0.28}$ & $0.26_{-0.03}^{+0.02}$ & $1.35_{-1.33}^{+0.09}$\\ 
        \\
        DS3 & $0.18_{-0.02}^{+0.02}$ & $0.61_{-0.07}^{+0.07}$ & $1.25_{-1.25}^{+1.06}$ & $6.85_{-3.03}^{+1.78}$ & $0.80_{-0.03}^{+0.20}$ & $0.73_{-0.04}^{+0.27}$ & $0.42_{-0.02}^{+0.02}$ & $0.82_{-0.79}^{+0.35}$ \\   
        \\
        DS4 & $0.19_{-0.03}^{+0.02}$ & $0.67_{-0.10}^{+0.04}$ & $3.04_{-3.04}^{+0.95}$ & $8.61_{-4.31}^{+1.16}$ & $0.84_{-0.01}^{+0.16}$ & $0.77_{-0.00}^{+0.23}$ & $0.63_{-0.58}^{+0.56}$ & $0.41_{-0.01}^{+0.02}$ \\
        \\
        \hline \hline
        \label{table:model_c_results}
   \end{tabular}
   \end{center}
\end{table*}

\subsection{Model D - Two-envelope model with reported pulsar density errors}

\begin{figure*}[!htbp]
\centering
\subfloat[DS3: Zeeman plus pulsar data.]{\includegraphics[scale=0.34]{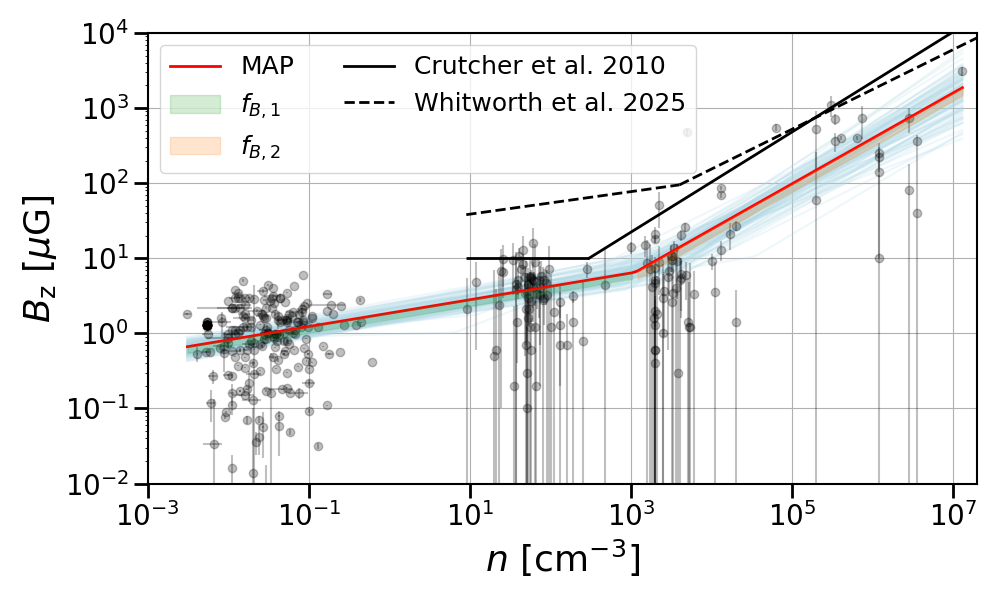}}
\subfloat[DS4: Zeeman plus pulsar, outlier removed.]{\includegraphics[scale=0.34]{figs/model_C/no_outliers/posterior_fit_.png}}\\
\caption{MAP results for model D, including explicit pulsar density errors where available. All notations are the same as Fig.~\ref{fig:model_C_grid}.}
\label{fig:model_D_grid}
\end{figure*}

Table~\ref{table:model_d_results} presents the MAP results for Model~D, which incorporates the reported errors on $n$ for the extended datasets. Figure~\ref{fig:model_D_grid} shows the MAP fits, with the two envelope scaling factors $f_{B,1}$ and $f_{B,2}$, as in Model~C.

In this model, $\alpha_1$ remains consistent with the previous models, while $\alpha_2$ is similar to Model~C, whilst in DS3 it is slightly increased and in DS4 it is slightly decreased. The break density $n_0$ increases from DS3 to DS4, by a factor of $2.38$. The inferred values for $B_0$ are also similar, with the value for DS4 being consistent with the original value of $10\,\mu$G in \citetalias{Crutcher2010}.

The two envelope scaling factors $f_{B,1}$ and $f_{B,2}$ are comparable when transitioning from diffuse ($f_{B,1}$) to dense ($f_{B,2}$) gas, with only a small decrease in the value when the outlier is removed. This suggests that the two-envelope approach is robust. The estimated $R_n$ is similar between the two datasets, though higher that the value previously reported in \citet{Crutcher2010}, giving an error factor of $\sim 4.5$--4.8. The intrinsic scatter parameter $\sigma_B$ shows little variation and in contrast to Model~C slightly decreases when the outlier is removed, reflecting the tighter density constraints in this model.

\begin{table*}[!htbp]
    \begin{center}
    \caption{MAP values from model~D }
    \begin{tabular}{lccccccccc}  
    \hline
    \\
        Data &  $\alpha_1$ & $\alpha_2$ & $n_0$ & $B_0(\mu{\rm G})$ & $f_{B,1}$ & $f_{B,2}$ & $\sigma_B$ & $R_n$\\
        Set & && $(10^3{\rm cm}^{-3})$ &&&&&
        \\
        \\
        \hline \hline 
        \\
        DS3 & $0.18_{-0.02}^{+0.03}$ & $0.62_{-0.07}^{+0.06}$ & $1.40_{-1.40}^{+0.73}$ & $6.97_{-3.28}^{+1.77}$ & $0.82_{-0.01}^{+0.18}$ & $0.81_{-0.04}^{+0.19}$ & $0.42_{-0.02}^{+0.02}$ & $0.65_{-0.64}^{+0.20}$ \\
        \\
        DS4 & $0.20_{-0.03}^{+0.01}$ & $0.65_{-0.07}^{+0.06}$ & $3.33_{-3.22}^{+0.27}$ & $10.11_{-5.97}^{+0.71}$ & $0.78_{-0.05}^{+0.22}$ & $0.78_{-0.01}^{+0.22}$ & $0.41_{-0.02}^{+0.02}$ & $0.68_{-0.67}^{+0.08}$ \\
        \\
        \hline \hline
   \end{tabular}
    \label{table:model_d_results}
    \end{center}
\end{table*}

\section{Discussion}
\label{sec:discussion}

\subsection{Model Uncertainties}
We have shown that switching from a single envelope scaling parameter $f_B$ across the full data, to using two, $f_{B,1}$ and $f_{B,2}$ on either side of the break density $n_0$, and including an intrinsic scatter term $\sigma_B$, enables the models to achieve much tighter posterior constraints. By allowing $f_{B,1}$ and $f_{B,2}$ to independently scale the upper envelope in the low- and high-density regimes, the model can more flexibly accommodate variations in magnetic field strength across different environments. The inclusion of $\sigma_B$ acknowledges that the reported observational uncertainties may underestimate the true scatter in $B$. As a result, the model does not rely solely on formal measurement errors, but also models intrinsic astrophysical variation, yielding MAP estimates that are less likely biased by the assumed under-predictions of LOS measurements. This increased complexity of the model leads to a more complete exploration of parameter space, allowing a more robust and reliable characterization of the $B$--$n$ relation.

The two envelope model also has the advantage that the MAP upper envelope becomes more tightly constrained, with decreased posterior uncertainty. However, this is accompanied by an increase in the intrinsic scatter parameter, $\sigma_B$. The implication of this is that, although the model is able to characterize the large-scale trend more accurately, it attributes more of the variance in the observed $B$ to intrinsic astrophysical scatter, rather than uncertainties in the envelope itself. As a result, the errors on individual predicted values of $B$ are not reduced, but may even grow, reflecting diversity in magnetic field strengths at a given density. In other words, the more robustly defined MAP line is achieved at the cost of acknowledging greater intrinsic dispersion among data points.

One of the key questions that has arisen for the $B$--$n$ relation is the implied errors on $n$. In the original Bayesian work by \citetalias{Crutcher2010} a fixed error of a factor of two (equivalent to $R_n = 0.30$) was used. In later work this was argued to be too small, with a value of nine being reported by \citet{Jiang2020}. One of the advantages of our new hierarchical Bayesian approach is the ability to infer a log-density intrinsic scatter $R_n$ directly from the data, similar to what was done by \citet{Jiang2020}. 

Our directly inferred $R_n$ values are consistent with those found in previous studies. Across our models and datasets, $R_n$ remains of order unity but is dataset-dependent and not uniformly well constrained.

If our model were simply masking density scatter with flexibility in other parameters, we would expect Model~B, where $R_n$ is fixed, to provide a better fit with more robust posteriors at all $R_n$. However, this is not the result; Model~B does not yield better-constrained or more physically plausible results. In fact, within Model~B, fixing $R_n$ does not yield uniformly improved or more physically plausible solutions across datasets, and the inferred break density $n_0$ remains sensitive to the chosen fixed value of $R_n$. This arises because a fixed $R_n$ controls the allowed density range, which can shift $n_0$ in order to accommodate the high-$n$ data. In contrast, our variable‑$R_n$ models A, C, and D recover $R_n$ values in the range 0.4--1.1, larger than the assumption of \citetalias{Crutcher2010}, but smaller than those used by \citetalias{Whitworth2025} or \citet{Jiang2020}. This indicates that the intrinsic scatter of density required by the present data is dependent on the model and how the data are treated. Part of this may reflect the heterogeneous nature of the sample itself, since the data lines of sight encompass clouds and diffuse regions that differ markedly in size, morphology, and environment, which our approach does not take into account.

Recall that $R_n$ describes the intrinsic scatter of $n$. In our Bayesian framework, it is largely constrained by data located near $n_0$, as points close to $n_0$ are more sensitive to the sharpness of the transition between the two regimes. The smaller the value of $R_n$ the more constrained the break density becomes. Thus, the population of measurements near $n_0$ has an influence on how well $R_n$ can be determined.

\subsection{Comparing models across datasets}
\label{sec:single_slope}
A key question that needs to be considered is whether increasing model flexibility improves genuine predictive performance or instead leads to overfitting. To test this, we compare models separately within each dataset using PSIS-LOO.

PSIS-LOO estimates expected out-of-sample predictive accuracy for Bayesian models without requiring a full refit for every leave-one-out (LOO) split. Standard importance-sampling LOO can become unstable when a few observations receive very large weights. PSIS stabilizes this calculation by smoothing the upper tail of the importance-weight distribution with a generalized Pareto fit, yielding more stable LOO estimates \citep[see][and references therein]{Vehtari2016}. We report $\mathrm{elpd}_{\mathrm{loo}}$ in Table \ref{tab:PSIS_LOO} along with its standard error. It predicts how well the model will fit unseen data; a higher (less negative) score means the Bayesian model has a better expected out-of-sample predictive fit.

Table~\ref{tab:pareto_k} reports Pareto-$k$ diagnostics to assess the reliability of the PSIS-LOO approximation. Here, $k_{\rm max}$ is the maximum Pareto shape parameter across observations, $k_{\mathrm{med}}$ is the median across observations, and $f_{>0.7}$ is the fraction of observations with $k>0.7$. Larger $k$ values indicate that the approximation is less reliable for those observations; thus, these diagnostics indicate approximation reliability and influential points, rather than overall model quality by themselves.

Quantitatively, the differences among Models A, B($R_n=0.3$), C, and D are modest in absolute terms for the Zeeman+pulsar sets, whereas fixing large density-scatter values ($R_n=0.7$ and $0.95$) consistently degrades the predictive fit. Whilst the Pareto-$k$ diagnostics reveal isolated high maximum values (e.g., $k_{\max} = 0.886$ for Model A in DS1 and $0.748$ for Model C in DS1), the negligible fraction of points above $0.7$ ($f_{>0.7} = 0.007$) and near-zero medians confirm that the PSIS-LOO approximations are overall robust, barring a few influential data points.

To test whether a broken power-law model is necessary, we also ran the PSIS-LOO on our preferred configuration (Model~D on DS4, see Section \ref{sec:opt} for details) and compared the two-slope formulation to a single-slope variant (D1) with the same likelihood and pulsar-error treatment (see Appendix~\ref{app:single_slope} for a full description of the model and MAP results). This comparison gives $\Delta\mathrm{elpd}_{\mathrm{loo}} = 27.02 \pm 7.15$ in favour of the two-slope model, corresponding to an approximate $\sim 3.8\sigma$ improvement, assuming Gaussian error propagation for $\Delta\mathrm{elpd}_{\mathrm{loo}}$. This indicates that the additional flexibility is supported by out-of-sample predictive performance rather than over-fitting.

\begin{table*}[ht]
\begin{center}
\caption{PSIS-LOO model comparison across data sets.}   
\label{tab:PSIS_LOO}
\begin{tabular}{lcccc}
\hline
Model & DS1 & DS2 & DS3 & DS4 \\
\hline
\\
A     & -114.02 $\pm$ 14.23          & -100.61 $\pm$ 10.52          & -270.61 $\pm$ 17.81         & -263.54 $\pm$ 17.51 \\
\\
B ($R_n = 0.3$) & \textbf{-112.58 $\pm$ 13.43} & \textbf{-100.46 $\pm$ 9.58}  & -270.84 $\pm$ 17.83         & -263.23 $\pm$ 17.56 \\
\\
B ($R_n = 0.7$) & -142.32 $\pm$ 6.62           & -138.71 $\pm$ 6.32           & -317.69 $\pm$ 9.44          & -314.23 $\pm$ 9.24  \\
\\
B ($R_n = 0.95$) & -166.59 $\pm$ 5.38           & -164.15 $\pm$ 5.29           & -383.27 $\pm$ 6.87          & -380.70 $\pm$ 6.79  \\
\\
C     & -114.12 $\pm$ 15.21          & -100.64 $\pm$ 10.81          & \textbf{-270.28 $\pm$ 18.53}        & -263.30 $\pm$ 18.16 \\
\\
D     & --                           & --                           & -270.36 $\pm$ 18.50 & \textbf{-262.97 $\pm$ 18.17} \\
\\
D1     & --                           & --                           & -- & -289.99 $\pm$ 16.71 \\
\\
\hline
\end{tabular}
\end{center}
   \par\noindent\textit{Notes.} 
   Values are $\mathrm{elpd}_{\mathrm{loo}} \pm \mathrm{SE}$; within each dataset column, larger (less negative) values indicate better expected out-of-sample predictive performance. Models with high density scatter ($R_n = 0.7$, $0.95$) and Model D1 have substantially lower scores than the other models.
\end{table*}

\begin{table*}[ht]
\begin{center}
\caption{
PSIS-LOO Pareto-$k$ diagnostics across datasets.}
\label{tab:pareto_k}
\begin{tabular}{l*{12}{c}}
\hline \\
\multirow{2}{*}{Model}
& \multicolumn{3}{c}{DS1}
& \multicolumn{3}{c}{DS2}
& \multicolumn{3}{c}{DS3}
& \multicolumn{3}{c}{DS4} \\
\cline{2-13}
\\
& $k_{\max}$ & $k_{\rm med}$ & $f_{>0.7}$
& $k_{\max}$ & $k_{\rm med}$ & $f_{>0.7}$
& $k_{\max}$ & $k_{\rm med}$ & $f_{>0.7}$
& $k_{\max}$ & $k_{\rm med}$ & $f_{>0.7}$ \\
\\
A
& 0.886 &  0.011 & 0.007
& 0.442 & -0.005 & 0.000
& 0.316 & -0.070 & 0.000
& 0.253 & -0.050 & 0.000 \\ \\

B ($R_n = 0.3$)
& 0.425 &  0.016 & 0.000
& 0.349 & -0.025 & 0.000
& 0.356 & -0.028 & 0.000
& 0.392 & -0.047 & 0.000 \\ \\

B ($R_n = 0.7$)
& 0.380 & -0.018 & 0.000
& 0.322 & -0.012 & 0.000
& 0.193 & -0.022 & 0.000
& 0.205 & -0.019 & 0.000 \\ \\

B ($R_n = 0.95$)
& 0.230 &  0.002 & 0.000
& 0.363 & -0.034 & 0.000
& 0.192 & -0.029 & 0.000
& 0.208 & -0.035 & 0.000 \\ \\

C
& 0.748 & -0.008 & 0.007
& 0.430 & -0.039 & 0.000
& 0.424 & -0.036 & 0.000
& 0.314 & -0.035 & 0.000 \\ \\

D
& \multicolumn{3}{c}{--}
& \multicolumn{3}{c}{--}
& 0.308 & -0.049 & 0.000
& 0.399 & -0.042 & 0.000 \\ \\

D1
& \multicolumn{3}{c}{--}
& \multicolumn{3}{c}{--}
& \multicolumn{3}{c}{--}
& 0.254 & -0.042 & 0.000 \\ \\
\hline
\end{tabular}
\end{center}
\par\noindent\textit{Notes.} 
Entries report the maximum Pareto shape parameter $k_{\max}$, the median Pareto shape parameter $\tilde{k}$, and the fraction $f_{>0.7}$ of observations with $k>0.7$. Lower values indicate more reliable PSIS-LOO approximations.
\end{table*}

\subsection{Our optimal model}
\label{sec:opt}
We show in Appendix \ref{ap:plots} that Models~C and D exhibit less curved parameter degeneracies and tighter joint posteriors compared to Models~A and B. Model~D in particular, which incorporates the errors reported on $n$ from the pulsar data of \citet{Seta2025}, provides well-constrained and unimodal posteriors and is preferred for DS4 in PSIS-LOO, so we adopt it as our benchmark model.

\begin{figure}
    \centering
    \includegraphics[scale=0.34]{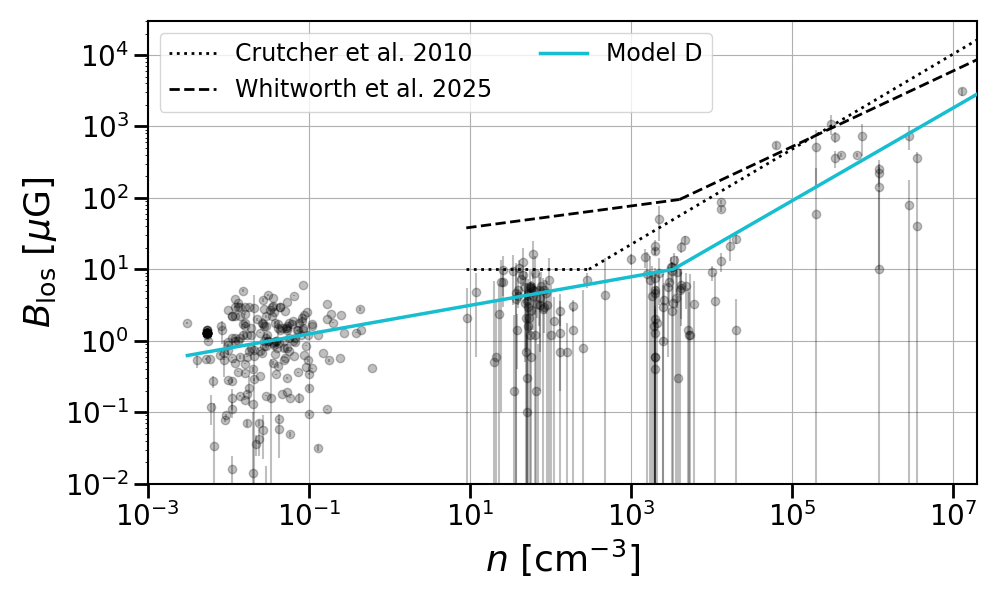}
    \caption{Our optimal MAP results from Model~D applied to DS4 showing the new relationship we present in this work in {\em cyan}, \protect\citetalias{Crutcher2010} upper-limit power law {\em (solid line)} and \protect \citetalias{Whitworth2025} MAP {\em (dashed line)}.}
    \label{fig:money}
\end{figure}

Examining the results for all models, it is clear that the new datasets extended to include pulsar measurements at low densities, DS3 and DS4, yield consistent trends for all error treatments and model variants. For the reasons discussed above, including the impact of the high-field outlier, we have determined that DS4 provides the best representation of the underlying $B$--$n$ relationship. We therefore adopt Model D applied to DS4 as our preferred reference model for the final $B$--$n$ relation. We have plotted this optimal model in Figure \ref{fig:money} compared to the data, and provided its numerical values in Table~\ref{table:model_d_results}.

\subsection{Physical Implications}
Our results reinforce those of \citetalias{Whitworth2025} that the magnetic field in the diffuse gas also shows variation with density, so the diffuse gas exponent $\alpha_1$ should be considered as a free parameter and not set to zero. Across all four models and datasets, seen in Tables \ref{table:model_a_results}--\ref{table:model_d_results}, we find that values of $\alpha_1$ cluster in a narrow range around $0.2$, spanning values of $0.17$--$0.22$ across the MAP fits. It is better constrained in the extended datasets including pulsar data, DS3 and DS4, across all models with much tighter uncertainties than in DS1 and DS2, consistent with the increased density of data at low $n$. This provides a narrower posterior for $\alpha_1$, reflecting increased certainty and reduced degeneracy with other parameters.

The high-density exponent $\alpha_2$ is broadly similar between models, with the clearest systematic change occurring when the outlier in Sgr B2 North is removed. This is understandable, as this data point has a high field strength, $B\,=\,460\,\mu$G, for its density of $n\,=\,5\,000\,\rm cm^{-3}$, which almost always lies above the break density $n_0$. Removing this point increases $\alpha_2$. With the outlier present, the model must accommodate its high field strength by flattening $\alpha_2$ to avoid over-predicting B for the remaining lower-field points at high density. When it is removed, these points can be fitted with a steeper, more physically consistent $\alpha_2$.

The break density $n_0$ lies in the range $1\,000$--$4\,000\,{\rm cm}^{-3}$, with higher values in some Zeeman-only fits. In Models A, C, and D, where the intrinsic scatter $R_n$ is inferred, we recover moderate values of $R_n \leq 1.0$ across all datasets but it can exceed unity in some model--dataset combinations. In Model~B where $R_n$ is fixed, the inferred break density remains sensitive to both the choice of $R_n$ and dataset. For example, DS4 has values for $n_0 = 2.08$, $2.10$, and $1.09\times10^3\,{\rm cm}^{-3}$ for fixed $R_n=0.30$, $0.70$, and $0.95$, respectively. These shifts reflect model assumptions about density-scatter treatment rather than a fundamental physical discrepancy.

Our results from Models~A, C, and D are broadly consistent with \citetalias{Whitworth2025} where $n_0\,=\, 4.0^{13.0}_{-3.0}\times10^3 \, \rm cm^{-3}$ is reported. As also noted in that paper, this break occurs around the point where the data becomes sparse at high densities $n = 10^4$--$ 10^7 \rm cm^{-3}$.

Physically, this transition could mark a regime in which the interplay between gravity, turbulence, and magnetic fields qualitatively changes. Thus, our results suggest that it is probably not a single sharp transition but a change over a range of transition densities. Such an idea has been proposed analytically. For example, \cite{Zhao2024_b} derive that results in a single continuous $B$--$n$ relationship. However, fitting such a model introduces new parameters and requires the development of a new Bayesian framework (probably first testing it over a range of numerical simulations as done for the broken power-law model in \citetalias{Whitworth2025}). Fitting such continuous-transition form (e.g., curved or exponential variants) is beyond the scope of this work. 

Recent numerical studies have suggested that at such high densities, the kinetic and turbulent energy, possibly driven by gravity, dominates over magnetic energy \citep{Ibanez2022,Li2025,McGuiness2025,Brucy2025}. As clouds collapse to form dense cores and stars, turbulent flows amplify the magnetic field through compression of field lines and small-scale dynamo action \citep{Seta2020}. However, the density scales and specific mechanisms at play are sensitive to environmental conditions such as cloud–cloud collisions and stellar feedback. Observational constraints on $n_0$ will provide key tests for theoretical models of the ISM and its magnetic and turbulent energy balance.

Previous work has argued that the value $\alpha_2 \simeq 0.66$ \citetext{\citetalias{Crutcher2010}; \citealt{Li2015,Jiang2020}} reflects a dynamical rather than purely kinematic collapse \citep{Mestel1966}. Our new models generally support this interpretation, although our inferred $\alpha_2$ values are slightly lower and may hint at the importance of the magnetic field in slowing but not preventing dynamical collapse.

The magnetic field plays a crucial role in the formation and shaping of molecular clouds, providing magnetic support that likely increases the mass and size of the cloud \citep{Girichidis2018,Whitworth2023, Robinson2023,Pillsworth2025a}. However, at higher densities $n > 10^4 \mbox{ cm}^{-3}$ where gravitational collapse dominates, the field becomes dynamically less important relative to gravity and turbulence, whose kinetic energy may be driven by colliding flows \citep{Brucy2025}, self-gravity \citep{Vazquez2019,Ibanez2022}, or stellar feedback \citep{Padoan2016}.

\subsection{Future extensions of observational data sets}

\begin{figure*}
    \centering
    \includegraphics[scale=0.2]{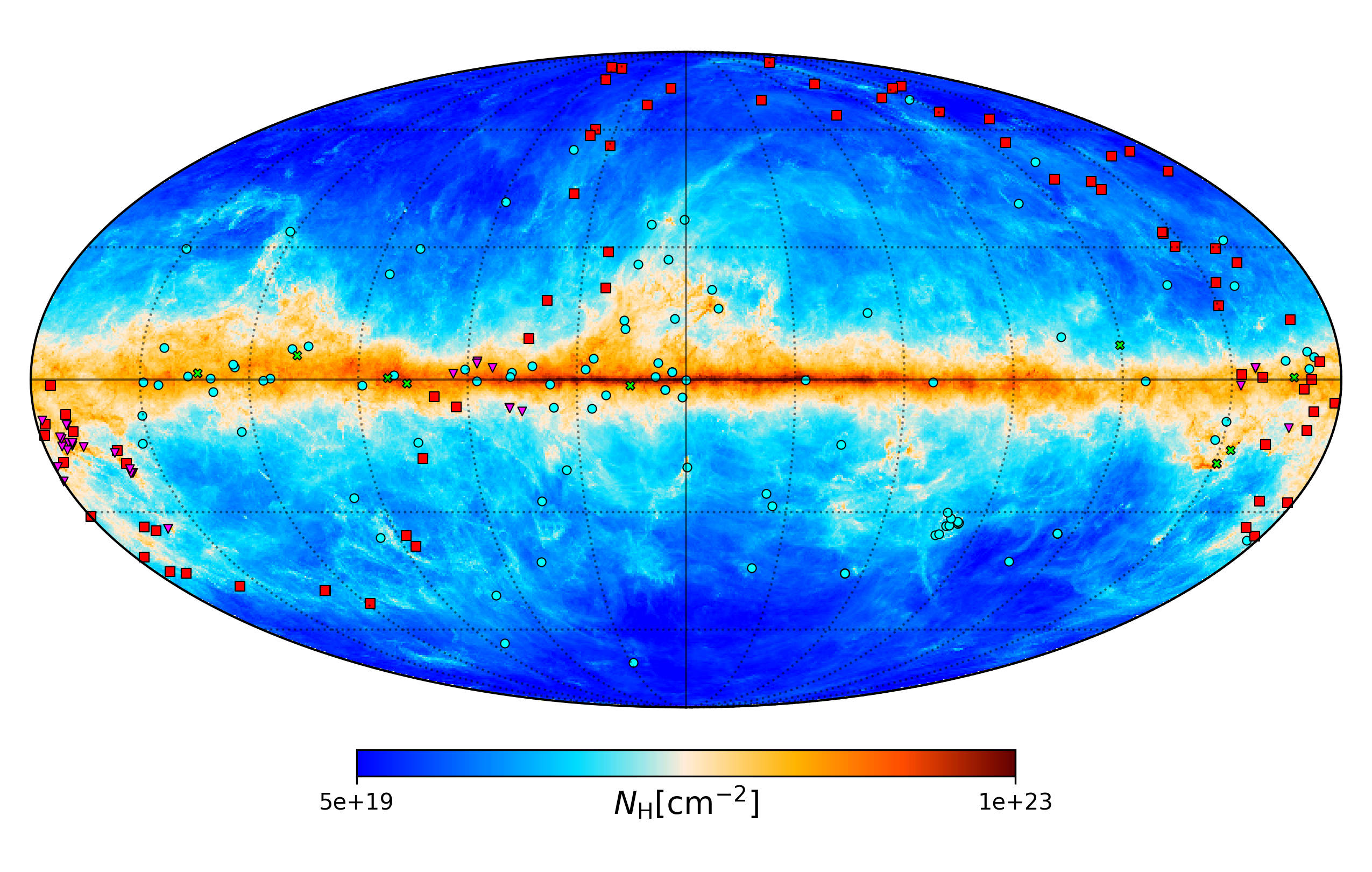}
    \caption{The position of observations used in this work overlaid on a Mollweide projection of dust column density derived from the {\it Planck} satellite observations \protect \citep{Plank2016,Planck2020}. The pulsars in \citet{Seta2025} are shown as {\em cyan circles}. The OH Zeeman measurements from \protect \citet{Troland2008} are shown as {\em magenta triangles}. The CN Zeeman measurements from \protect \citet{Falgarone2008} are {\em green crosses} and the H{\sc{i}} Zeeman measurements from \protect \citet{Heiles2004} are shown as {\em red squares}. It can be seen that there is a lack of coverage of Quadrant 4 of the galactic plane in the observations.}
    \label{fig:planck}
\end{figure*}

Having used the pulsar data to secure a more precise value for $\alpha_1$, further improvements to our results can be made by obtaining more observational data in the higher density regime. In Figure \ref{fig:planck}, we overlay all the observational data used in this work (both Zeeman and pulsar) on the Planck all-sky map of galactic dust density. Data coverage across Quadrant 4 of the galactic plane is rather sparse.\footnote{For the pulsar data coverage, see also Fig.~2 in \citet{Dhakal2025}.} Although the combined pulsar and Zeeman data provide strong constraints on the diffuse component of the $B$--$n$ relation, only 71 of the 347 data points probe regions with $n_0 > 300\rm \, cm^{-3}$, and 40 above our new break density of $n_0 = 3330\rm \, cm^{-3}$, 
tracing sightlines within or near the Galactic mid-plane. Expanding coverage into parts of the disk where the warm diffuse medium transitions to cold molecular gas would allow more direct constraints on $n_0$. Also, additional data in the dense regime would improve constraints on $\alpha_2$.

\section{CONCLUSIONS} 
\label{sec:conclusions}

In this work, we extended the $B$--$n$ relation published in \citetalias{Whitworth2025} by incorporating a large set of pulsar observations \citep{Seta2025}. This addition has enabled us to obtain new constraints on the exponent in the diffuse gas regime, with all our models and data sets recovering a non-zero power-law slope of $\alpha_1 \simeq 0.2$. The success of this approach stems from both the inclusion of pulsar data from \citet{Seta2025}, as well as a more rigorous treatment of density uncertainties through our hierarchical Bayesian framework.

Our most reliable MAP results come from Model~D with Data Set 4, which combines the Zeeman measurements reported in \citetalias{Crutcher2010} with new Zeeman data from \citet{Hwang2024} and the pulsar data, excluding the known Zeeman outlier in Sgr B2 North. The model includes reported uncertainties on $n$ for the low-density pulsars and allows for intrinsic variance in both $n$ and $B$. We note that the pulsar data measure $n_e$ whilst H\textsc{i}, OH and CN measure $n$, which should be kept in mind when interpreting the combined relation.

To summarize, our key results and conclusions are:
\begin{itemize}
    \item We fit a two-part power-law to the original Zeeman data of \citetalias{Crutcher2010} and a new extended data set that includes low-density pulsar measurements from \citet{Seta2025}, as well as extra Zeeman measurements from \citet{Hwang2024}.
    \item We developed four new hierarchical Bayesian models to examine the relationship between $n$ and $B$. We find that incorporating reported errors on $n$ from the pulsar data, allowing for intrinsic scatter in $n$ via a hyperparameter $R_n$, and modelling the intrinsic scatter in $B$ to account for geometric and observational uncertainties produces the most statistically robust results. 
    \item The inclusion of the pulsar data in the expanded hierarchical model (Model~D, DS4) yields the most robust observational $B$--$n$ relation, shown in Figure~\ref{fig:money}, with MAP parameters:
    \item[] $\alpha_1 = 0.20^{+0.01}_{-0.03}$, $\alpha_2 = 0.65^{+0.06}_{-0.07}$,
    \item[] $n_0 = 3330^{+270}_{-3220}$\,cm$^{-3}$, $B_0 = 10.11^{+0.71}_{-5.97}$\,$\mu$G,
    \item[] $f_{B,1} = 0.78^{+0.23}_{-0.05}$, $f_{B,2} = 0.78^{+0.22}_{-0.01}$,
    \item[] $\sigma_B = 0.41^{+0.02}_{-0.02}$, and $R_n = 0.68^{+0.08}_{-0.67}$.
\end{itemize}

The variance in $n_0$ may reflect the dynamical and anisotropic nature of the ISM, and that the magnetic field strength is tied to factors beyond just the transition from diffuse to dense gas. Although the diffuse component of the $B$--$n$ relation is now well constrained, only 40 of our 347 data points probe the dense gas. Looking ahead, expanding the observational coverage into regions where the warm diffuse medium transitions to cold molecular gas would sharpen constraints on $n_0$ and the slopes and advance our understanding of magnetic field dynamics across the multiphase ISM.

\section*{Acknowledgments}

The authors thank the anonymous referee for insightful comments and suggestions that helped to improve the clarity of our discussion, Sundar Srinivasan and Gwendoline Eadie for the original discussions on the implementation of hierarchical Bayesian techniques, and Chris McKee, Edith Falgarone, and Katia Ferri\`ere for interesting and detailed discussions on the results of \citetalias{Whitworth2025}. DW acknowledges support from the ANR CASCADE grant (ANR-24-ERCS-0004). AS is supported by the Australian Research Council through the Discovery Early Career Researcher Award (DECRA) Fellowship (project~DE250100003) funded by the Australian Government and the Australia-Germany Joint Research Cooperation Scheme of Universities Australia (UA--DAAD, 2025--2026). REP is supported by a Discovery Grant from the Natural Sciences and Engineering Research Council (NSERC) of Canada. M-MML acknowledges partial support from US NSF grant AST23-07950. JDS acknowledges financial support from the Austrian Science Fund (Fonds zur Förderung der wissenschaftlichen Forschung, FWF) through the ``Neutral Atomic Hydrogen in the solar neighborhood'' (NeAtHood) project (Grant DOI 10.55776/PAT6169824). AP acknowledges financial support from the UNAM-PAPIIT IN120226 grant, and the Sistema Nacional de Investigadores of SECIHTI, M\'exico. RSK acknowledges financial support from the ERC via Synergy Grant ``ECOGAL'' (project ID 855130) and from the German Excellence Strategy via the Heidelberg Cluster ``STRUCTURES'' (EXC 2181 - 390900948). In addition, RSK is grateful for funding from the German Ministry for Economic Affairs and Climate Action in project ``MAINN'' (funding ID 50OO2206), and from DFG and ANR for project ``STARCLUSTERS'' (funding ID KL 1358/22-1). 

\section*{Data Availability}

The hierarchical Bayesian analysis codes can be found at: \href{https://github.com/dwhtwrth-astro/hierarchical_Bayesian_2026.git}{\url{https://github.com/dwhtwrth-astro/hierarchical_Bayesian_2026.git}} whilst the data can be found in either the relevant papers or through contacting their authors.

\bibliographystyle{mnras}
\bibliography{paper} 

\begin{appendix}

\section{Appendix A}
\label{Append:code}

Whilst developing the hierarchical Bayesian models used in this work, we identified and corrected an implementation error in the original code used by \citetalias{Whitworth2025} that affected the inferred value of $B_0$. Specifically, a duplicate prior sampling statement on $\sigma_y$ in the HMC model overweighted the prior distribution on intrinsic scatter, which contributed to biasing the inferred value of $B_0$. Correcting this error motivated further improvements to the statistical framework. Following the approach of \citet{Jiang2020}, we introduced an envelope scaling factor $f_B$ to account for projection effects and an intrinsic scatter hyperparameter $f_n$ for $n$. 

\begin{figure}
    \centering
    \includegraphics[scale=0.34]{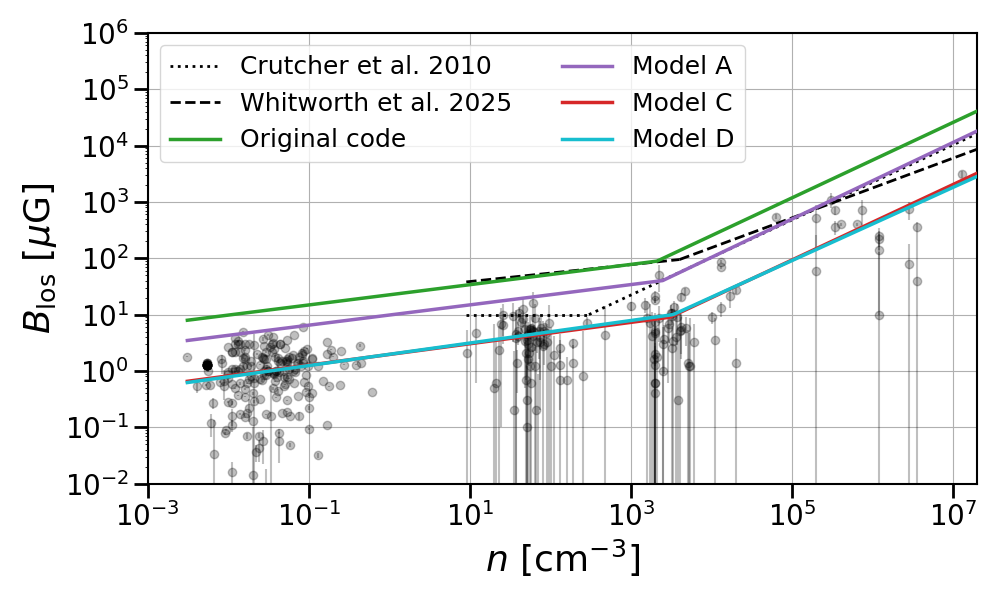}
    \caption{A comparison between the code from \protect\citetalias{Whitworth2025}, applied to DS4, and the code used in this paper for Models~A, C, and D. We see that the original incorrect code overpredicts the value for $B_0$.}
    \label{fig:compare}
\end{figure}

To assess the impact of these combined changes, we tested the extended Zeeman plus pulsar dataset DS4 using both the original code and our new implementation. Table~\ref{table:code_error} shows the results from this test compared to the results for Models~A, C, and D, and plotted in Figure~\ref{fig:compare}. The comparison reveals that $B_0$ is systematically overpredicted in the original code by factors of $\sim$2--10 depending on the model, whilst the power-law indices $\alpha_1$ and $\alpha_2$ and density break-point $n_0$ remain relatively stable across implementations.

The physical significance becomes evident when examining our more sophisticated Models~C and D. Using the corrected implementation and improved modelling, $B_0$ converges to $\sim 8\,\mu$G, close to the Milky Way's average magnetic field strength and the originally reported value of $10 \,\mu$G in \citetalias{Crutcher2010}. This convergence demonstrates that our improved hierarchical framework more accurately captures the underlying $B$--$n$ relationship.

\begin{table}
    \begin{center}
    \caption{Comparison between results in \protect\citetalias{Whitworth2025} and new models}
    \begin{tabular}{lcccc}  
    \hline
    \\
        & $\alpha_1$ & $\alpha_2$ & $n_0$, & $B_0,(\mu{\rm G})$ \\
        &&& $(10^3{\rm cm}^{-3})$ & \\
        \\
        \hline \hline 
        \\
        \citetalias{Whitworth2025} & $0.19_{-0.02}^{+0.02}$ & $0.67_{-0.07}^{+0.08}$ & $1.10_{-0.80}^{+2.00}$ & $81.0_{-22.0}^{+33.0}$ \\
        \\
        Model A & $0.18_{-0.02}^{+0.02}$ & $0.68_{-0.11}^{+0.04}$ & $2.45_{-2.45}^{+1.35}$ & $39.90_{-36.60}^{+7.80}$ \\
        \\
        Model C & $0.19_{-0.03}^{+0.02}$ & $0.67_{-0.10}^{+0.04}$ & $3.04_{-3.04}^{+0.95}$ & $8.61_{-4.31}^{+1.16}$ \\
        \\
        Model D & $0.20_{-0.03}^{+0.01}$ & $0.65_{-0.07}^{+0.06}$ & $3.33_{-3.22}^{+0.27}$ & $10.11_{-5.97}^{+0.71}$ \\
         \\
        \hline \hline
        \label{table:code_error}
   \end{tabular}
   \end{center}
   \textit{Notes.} A comparison between the code from \protect\citetalias{Whitworth2025} and the revised code used in this paper for Models~A, C, and D, all applied to DS4.
\end{table}

\section{Appendix B}
\label{ap:plots}

\begin{figure*}[!ht]
    \centering
    \includegraphics[scale=0.45]{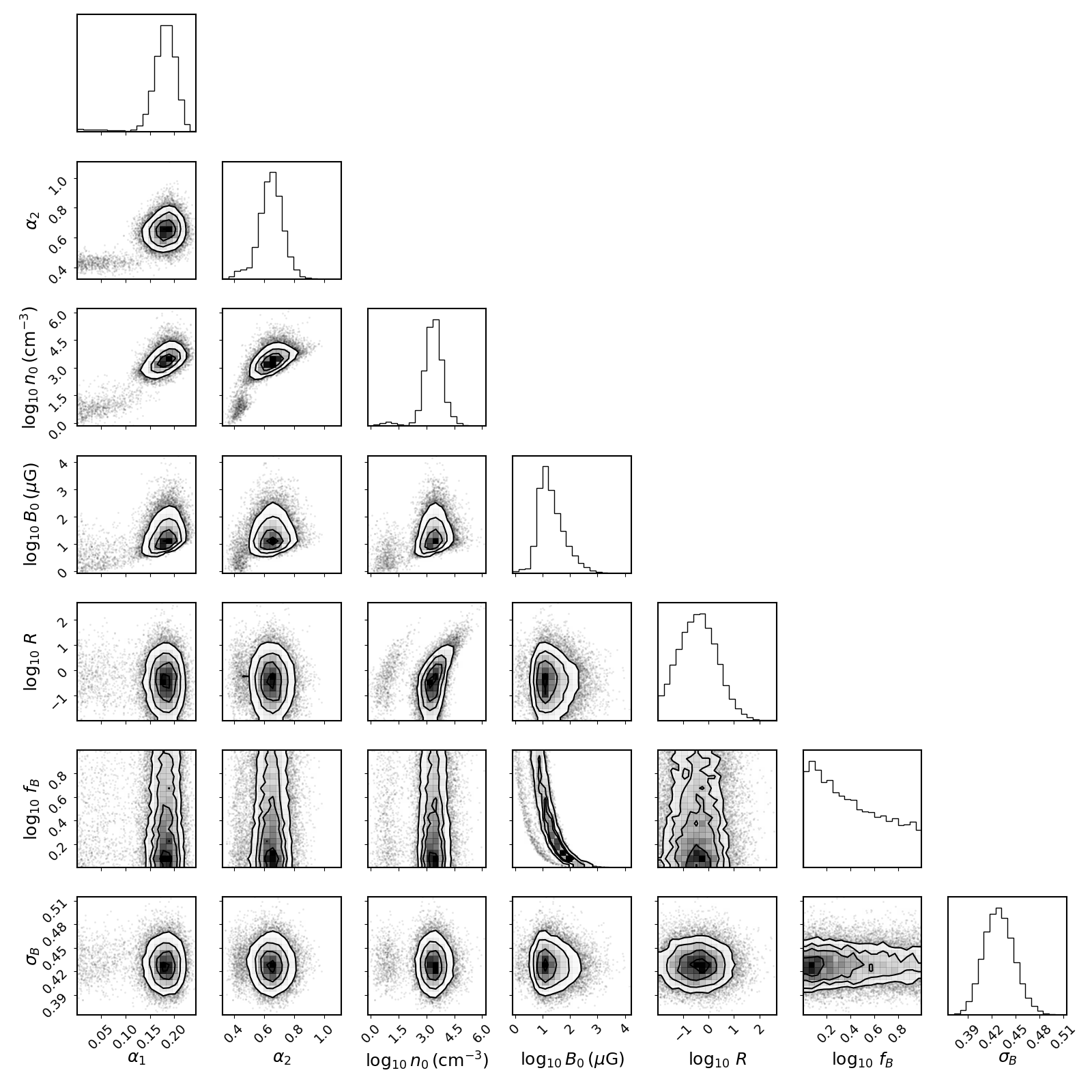}
    \caption{Corner plot for Model~A applied to DS4}
    \label{fig:corner_model_A}
\end{figure*}

\begin{figure*}[!ht]
    \centering
    \includegraphics[scale=0.5]{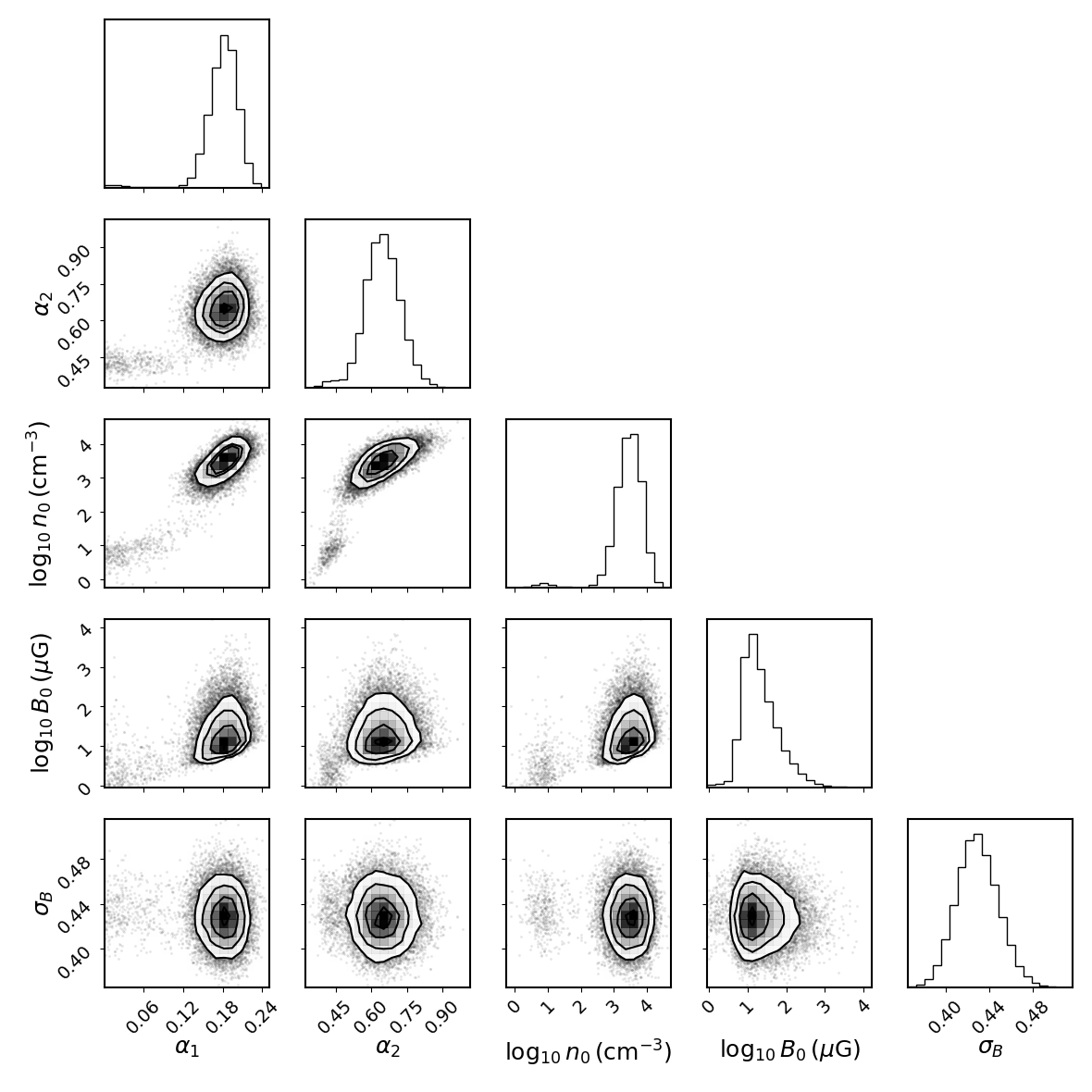}
    \caption{Corner plot for $R_n = 2$ for Model~B applied to DS4}
    \label{fig:corner_model_B}
\end{figure*}

\begin{figure*}[!ht]
    \centering
    \includegraphics[scale=0.4]{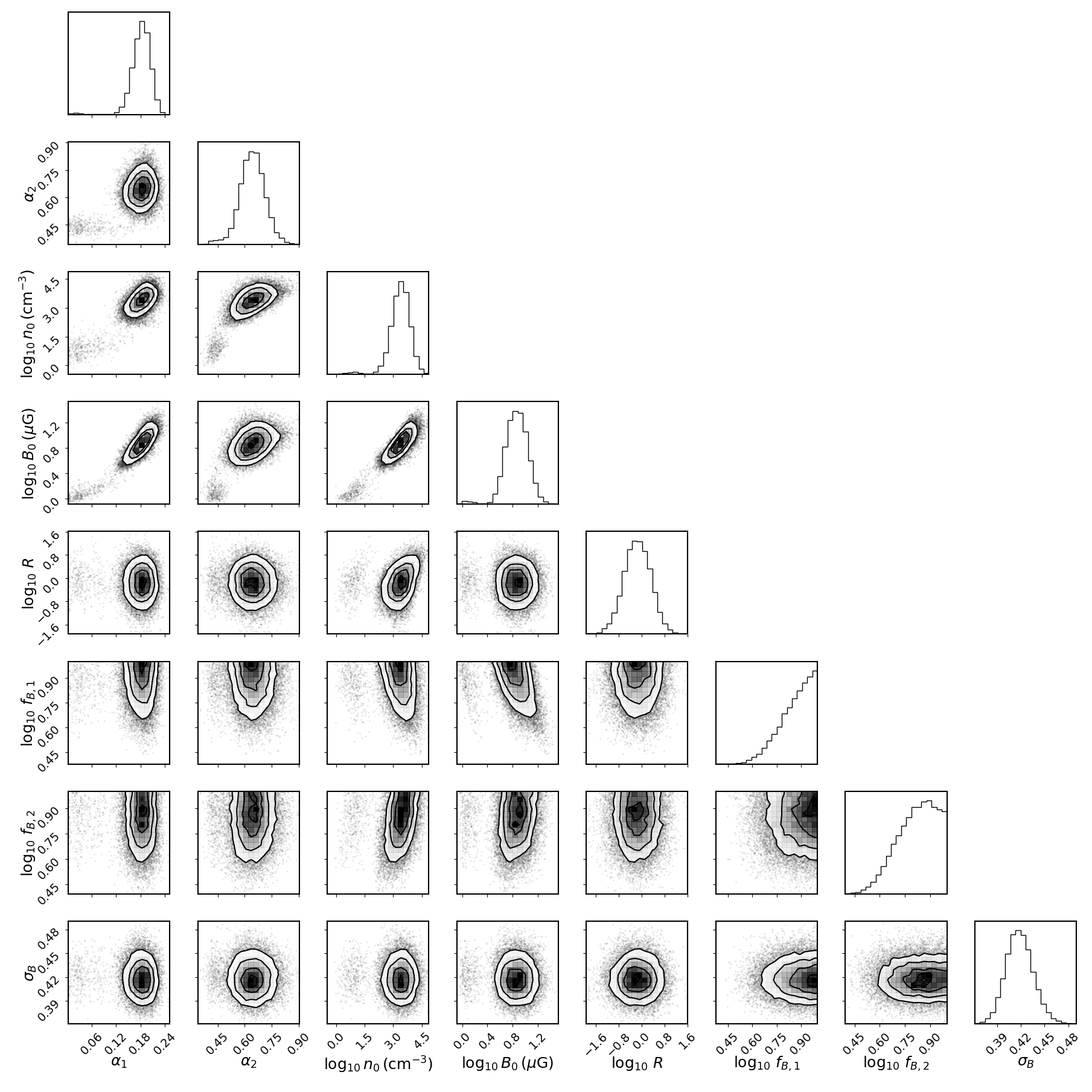}
    \caption{Corner plot for Model~C applied to DS4}
    \label{fig:corner_model_C}
\end{figure*}

\begin{figure*}[!ht]
    \centering
    \includegraphics[scale=0.4]{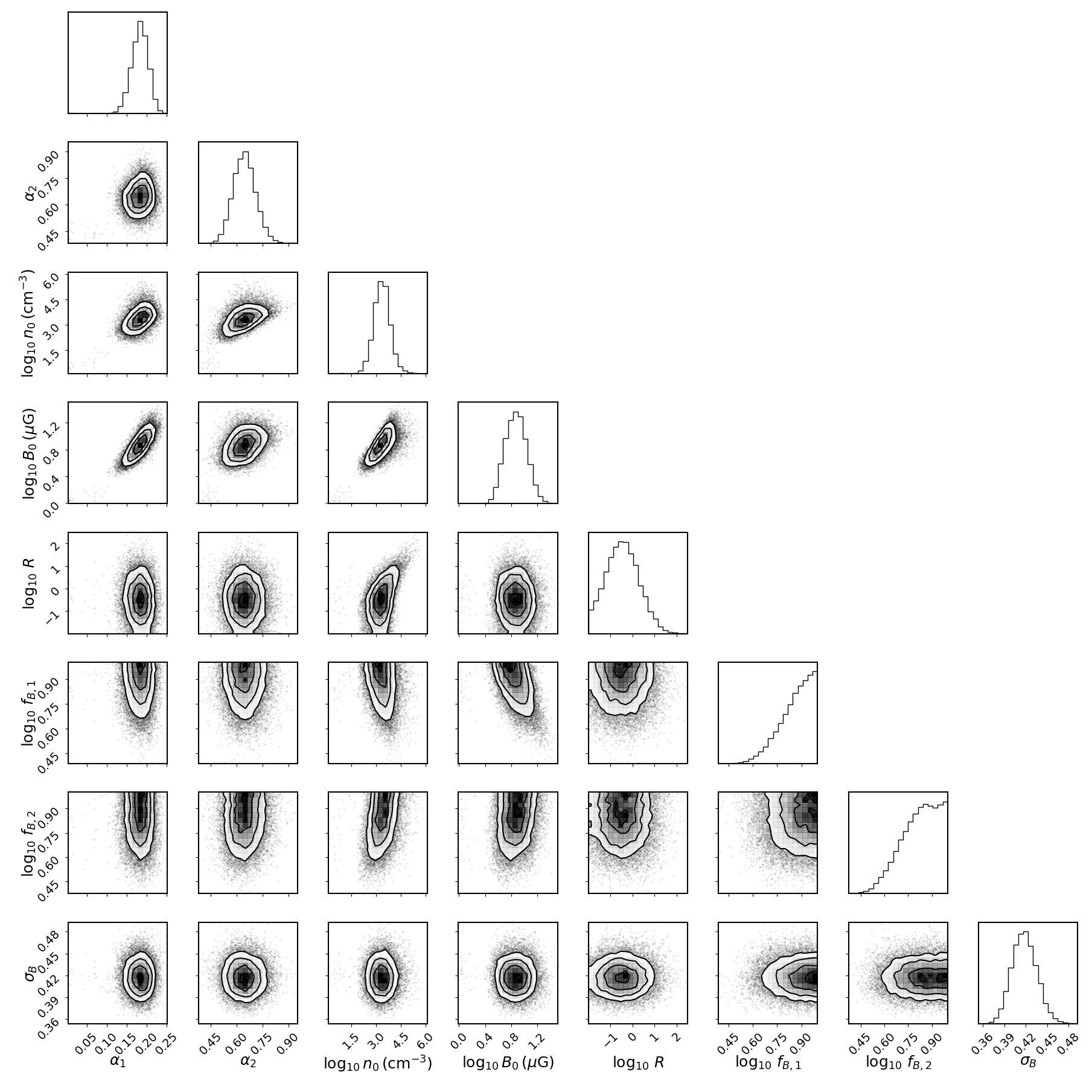}
    \caption{Corner plot for Model~D applied to DS4}
    \label{fig:corner_model_D}
\end{figure*}

Here we present the corner plots for DS4 for our models. For Model B, we show only the case with $R=2$. Models~A and B (Figures \ref{fig:corner_model_A} and \ref{fig:corner_model_B}) exhibit broader and more skewed posteriors, with mild hints of bi-modality, particularly in log$_{10}\,n_0$, where a small secondary peak is visible, and possibly in $\alpha_1$. Model~C (Figure \ref{fig:corner_model_C}) shows nearly Gaussian joint distributions, though weak bi-modality remains in log$_{10}$\,$n_0$ and log$_{10}$\,$B_0$, having mostly disappeared for $\alpha_1$. Model~D (Figure \ref{fig:corner_model_D}) exhibits unimodal, approximately Gaussian posteriors for all parameters, with only slight elliptical covariance and no strong degeneracies.

The sharp upper limits seen in $f_B$, $f_{B,1}$, and $f_{B,2}$ for Models~A, C, and D arise from how this parameter is implemented in the Bayesian framework. The prior on each is defined over log$_{10}$\,$f_B$ between 0 and 1 (see Table \ref{table:all_models}), and because it enters the model as an additive term in Equation \eqref{eq:model_A_field}, the allowed range enforces $f_B \leq 1$. In Models~C and D, where separate scaling factors $f_{B,1}$ and $f_{B,2}$ are used above and below the break density $n_0$, both parameters tend to saturate at the upper bound of 1, indicating that the data prefer minimal suppression of $B$ relative to $B_0$ in both regimes. In contrast, Model~A employs a single scaling factor over the full dataset, and its posterior for $f_B$ is correspondingly broader, reflecting the need to accommodate the full density range within one scaling parameter supporting our choice of Model D for our most robust result.

\section{Appendix C}
\label{ap:sensitivity}

In this Appendix, we examine the effects of Zeeman observational sensitivity using a simple diagnostic test. The aim is not to provide a complete treatment of selection effects, but rather to illustrate how an imposed sensitivity threshold can influence the inferred $B$--$\rho$ relation.The measurements are subject to a sensitivity limit set by the S/N. \cite{Troland2008} estimate that an S/N of $\sim 2.4$ corresponds to a field strength of $\sim 2.5\,\mu$G. In practice, this means that fields below $\sim 2.5\,\mu$G are increasingly difficult to distinguish from noise and may be reported as non-detections. Whilst \citetalias{Crutcher2010} took into account an upper limit on the magnetic field strength into account in their Bayesian analysis, the observational sensitivity effect was not explicitly accounted for; all available data were included in the analysis. We follow the same approach as \citetalias{Crutcher2010} for out four main data sets. However, it is prudent to test how a sensitivity threshold will affect the inferred parameters of the $B$--$\rho$ relation.

There are several approaches to determining how sensitivity limits affect Bayesian outcomes. Rigorous approaches caution against simply eliminating measurements that achieve only upper limits \citep{Loredo2004}. That said, this involves more detailed analysis that is beyond the scope of this work.

\begin{figure}
    \centering
    \includegraphics[scale=0.35]{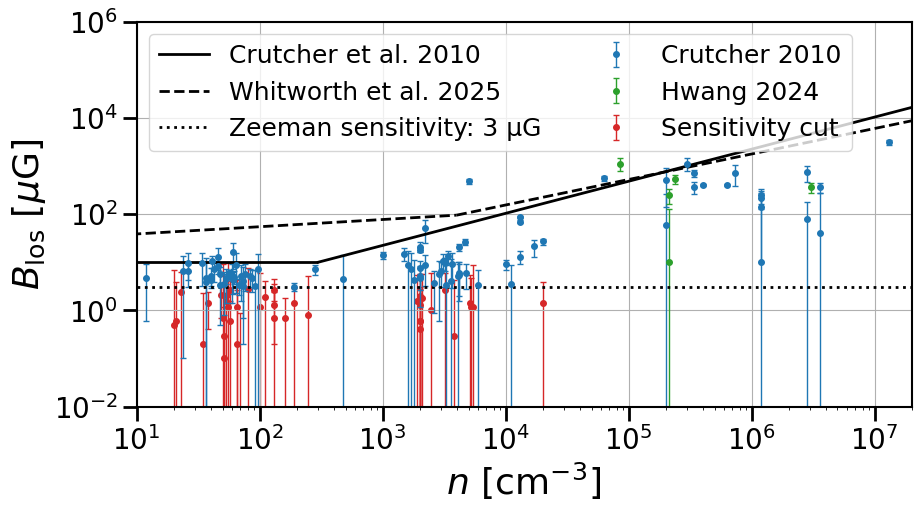}
    \caption{Zeeman data from \protect \citetalias{Crutcher2010} and \protect \cite{Hwang2024}. Lines show \protect\citetalias{Crutcher2010} upper-limit power law {\em (solid line)} and the \protect \citetalias{Whitworth2025} MAP result {\em (dashed line)}. We show in red the data cut due to the sensitivity limit we impose on the Zeeman data with the {\em dotted line} showing the limit across the plot.}
    \label{fig:Zeeman_sensitivity}
\end{figure}

We can nevertheless gain some preliminary understanding of such effects by adopting the conservative sensitivity limit of $3\,\mu$G and removing all Zeeman measurements below this value. This excludes 41 data points from the Zeeman sample. We then generate 2 additional data sets, one for the Zeeman only data, DS5, and one for the extended Zeeman plus pulsar data, DS6, removing the high-field outlier as well. Figure~\ref{fig:Zeeman_sensitivity} shows the adopted cut off (dotted line) and the excluded data in red.

\begin{itemize}
  \item DS5: As DS2 but with the sensitivity limit applied
  \item DS6: As DS4 but with the sensitivity limit applied
\end{itemize} 

We note, when the sensitivity limit cut is applied to Models C and D, the posterior can become sharply truncated, which can lead to excessively small step sizes or slow mixing during adaptation if a high target acceptance probability (\texttt{adapt\_delta}) is enforced. To maintain practical sampling efficiency and ensure robust exploration of challenging posterior regions, we lowered \texttt{adapt\_delta} to $0.85$ and increased the number of warm-up and sampling iterations to $10\,000$ each. This combination allows the sampler to traverse sharp features and boundaries more effectively, balancing a modest increase in the rate of divergent transitions with improved mixing and efficiency.

We emphasize that this is a deliberately simple diagnostic intended only to illustrate that sensitivity limits can affect the inferred fit. A more rigorous treatment would require a more elaborate Bayesian framework that explicitly models the selection effects, so we interpret these results with appropriate caution.

\begin{figure*}[!htbp]
\centering
\subfloat[Model~A DS5: DS2 with sensitivity limit applied.]{\includegraphics[scale=0.35]{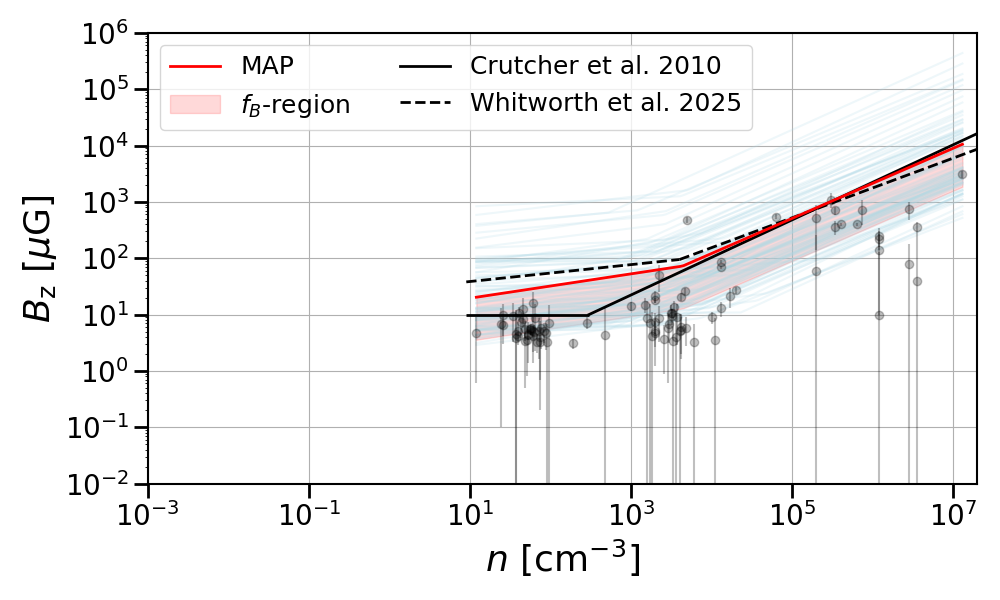}}
\subfloat[Model~A DS6: DS4 with sensitivity limit applied.]{\includegraphics[scale=0.35]{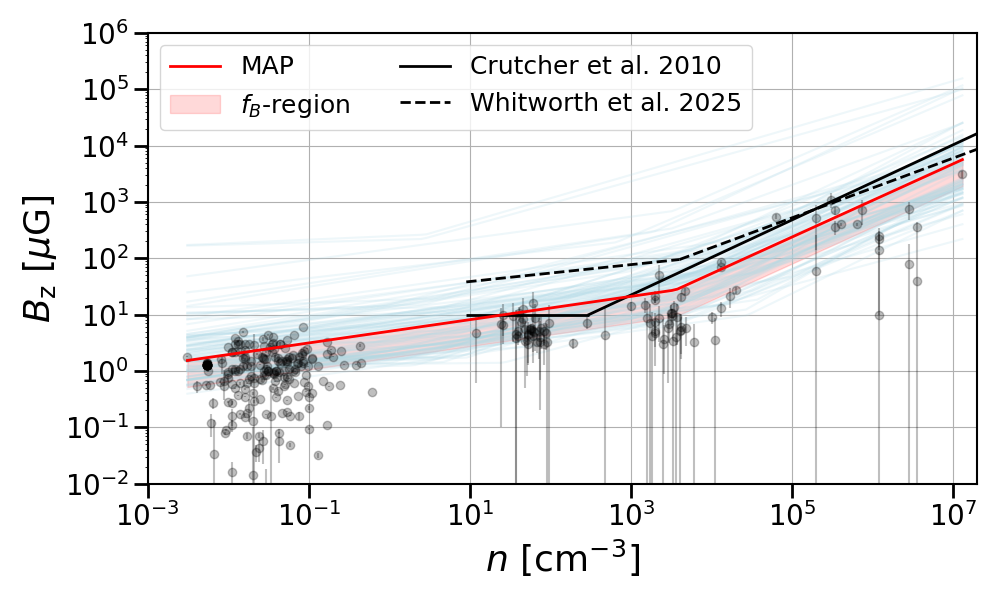}}\\
\subfloat[Model~B DS5: DS2 with sensitivity limit applied.]{\includegraphics[scale=0.35]{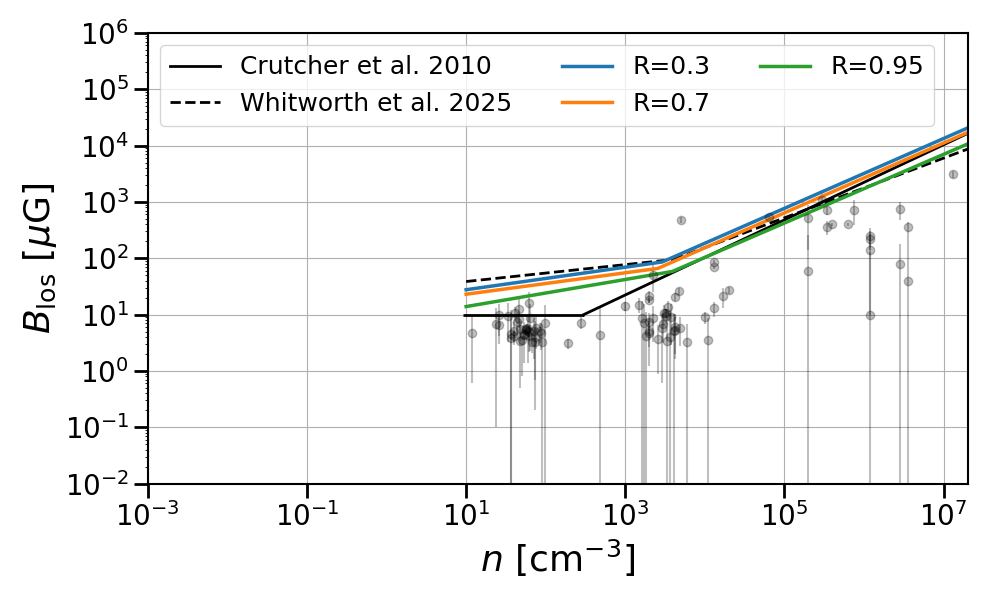}}
\subfloat[Model~B DS6: DS4 with sensitivity limit applied.]{\includegraphics[scale=0.35]{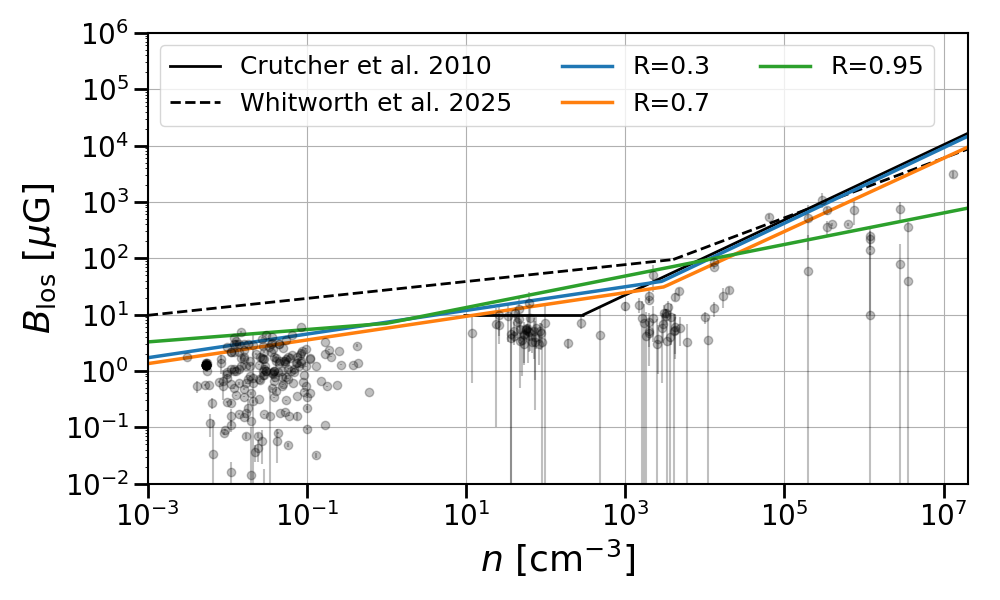}}\\
\subfloat[Model~C DS5: DS2 with sensitivity limit applied.]{\includegraphics[scale=0.35]{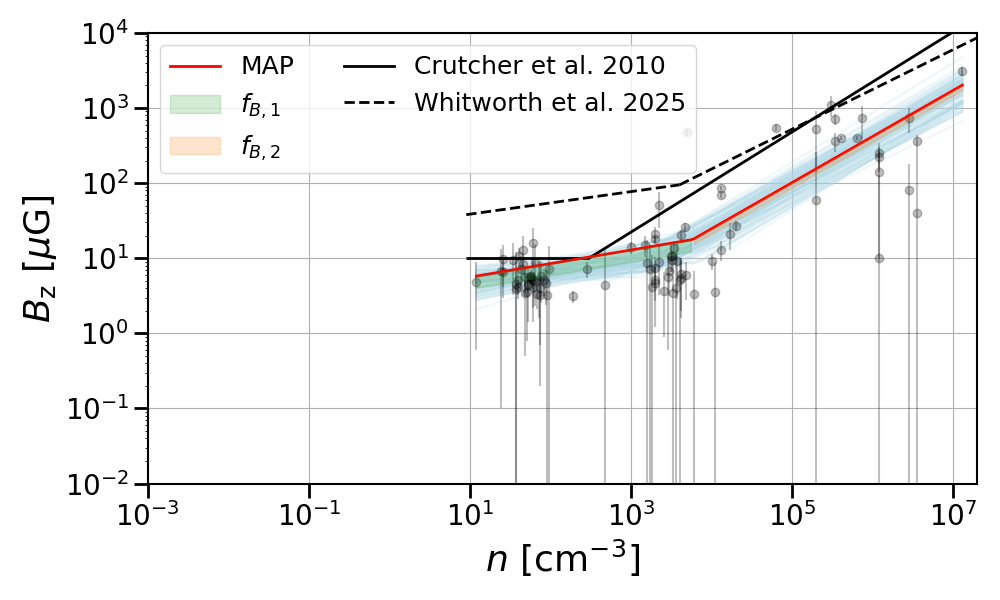}}
\subfloat[Model~C DS6: DS4 with sensitivity limit applied.]{\includegraphics[scale=0.35]{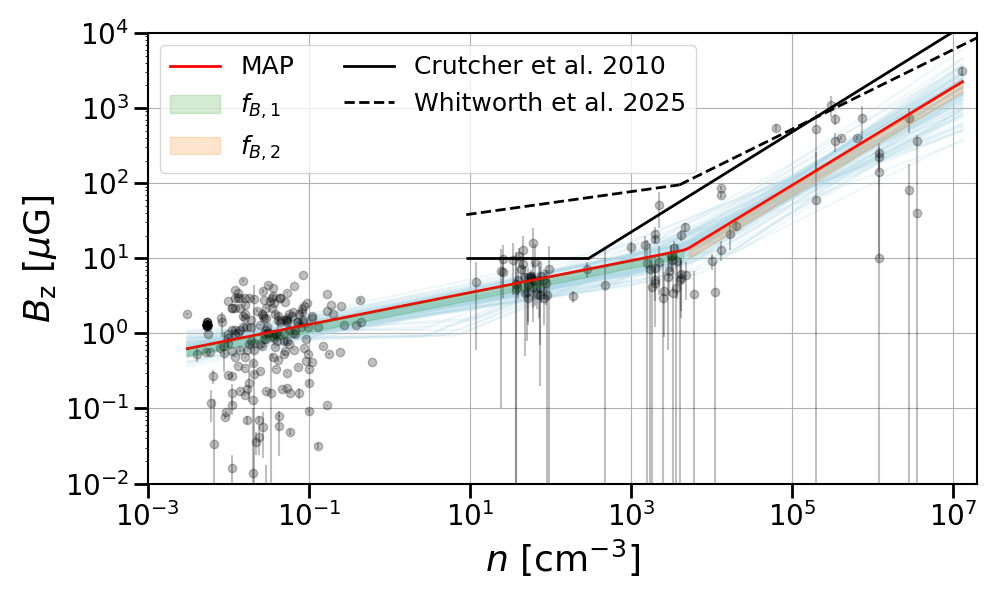}}\\
\subfloat[Model~D DS6: DS4 with sensitivity limit applied.]{\includegraphics[scale=0.35]{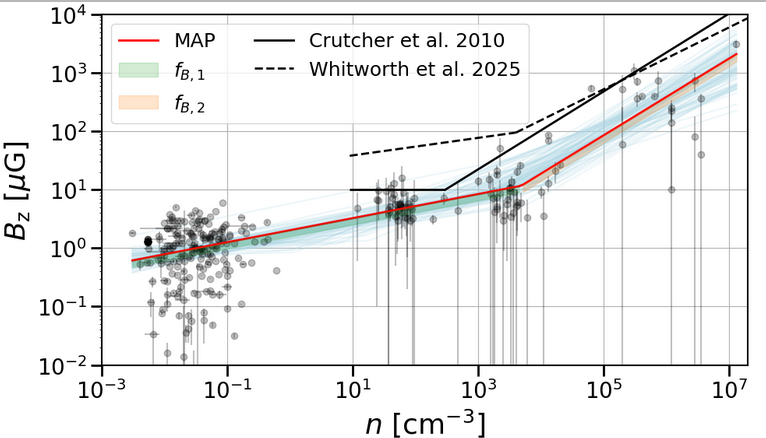}}
\caption{MAP results for all models when the sensitivity cut is included.}
\label{fig:sensitivity_limit}
\end{figure*}

Figure \ref{fig:sensitivity_limit} depicts the results for each model fitting to the new data sets. Tables \ref{table:sensitivity_results_model_A}, \ref{table:sensitivity_results_model_B}, \ref{table:sensitivity_results_model_C} and \ref{table:sensitivity_results_model_D} show the MAP results for the new data sets. The most notable change is in $n_0$ and $R_n$. In the Zeeman only data set DS5, $R_n$ decreases from the main models, whereas in DS6 it increases. The break density $n_0$ also shifts, either increasing or decreasing, depending on the model. This is expected because the sensitivity cut removes many of the low-field Zeeman measurements that lie close to the transition region and therefore provide the strongest constraint on the location of the break.

\begin{table*}[!htbp]
    \begin{center}
    \caption{MAP values from sensitivity limit analysis of model~A on the reduced data sets}
    \begin{tabular}{lccccccccc}  
    \hline
    \\
        Data &  $\alpha_1$ & $\alpha_2$ & $n_0$ & $B_0(\mu{\rm G})$ & $f_{B}$ & $\sigma_B$ & $R_n$\\
        Set & & & $(10^3{\rm cm}^{-3})$ &&&&&
        \\
        \\
        \hline \hline
        \\      
        DS5 & $0.22_{-0.20}^{+0.03}$ & $0.62_{-0.07}^{+0.04}$ & $4.36_{-4.03}^{+0.50}$ & $72.71_{-67.39}^{+8.44}$ & $0.18_{-0.17}^{+0.28}$ & $0.31_{-0.02}^{+0.04}$ & $0.53_{-0.52}^{+0.92}$ \\   
        \\
        DS6 & $0.21_{-0.06}^{+0.04}$ & $0.65_{-0.25}^{+0.01}$ & $3.58_{-3.58}^{+0.29}$ & $27.46_{-26.61}^{+0.19}$ & $0.34_{-0.33}^{+0.20}$ & $0.43_{-0.02}^{+0.02}$ & $0.57_{-0.56}^{+0.39}$ \\
        \\
        \hline \hline
   \end{tabular}
    \label{table:sensitivity_results_model_A}
    \end{center}
    \par\noindent\textit{Notes.} 
    The same headings and labels as Table \ref{table:model_a_results} except we have used the reduced data sets as described in the text above.
\end{table*}

\begin{figure*}[!ht]
    \centering
    \includegraphics[scale=0.4]{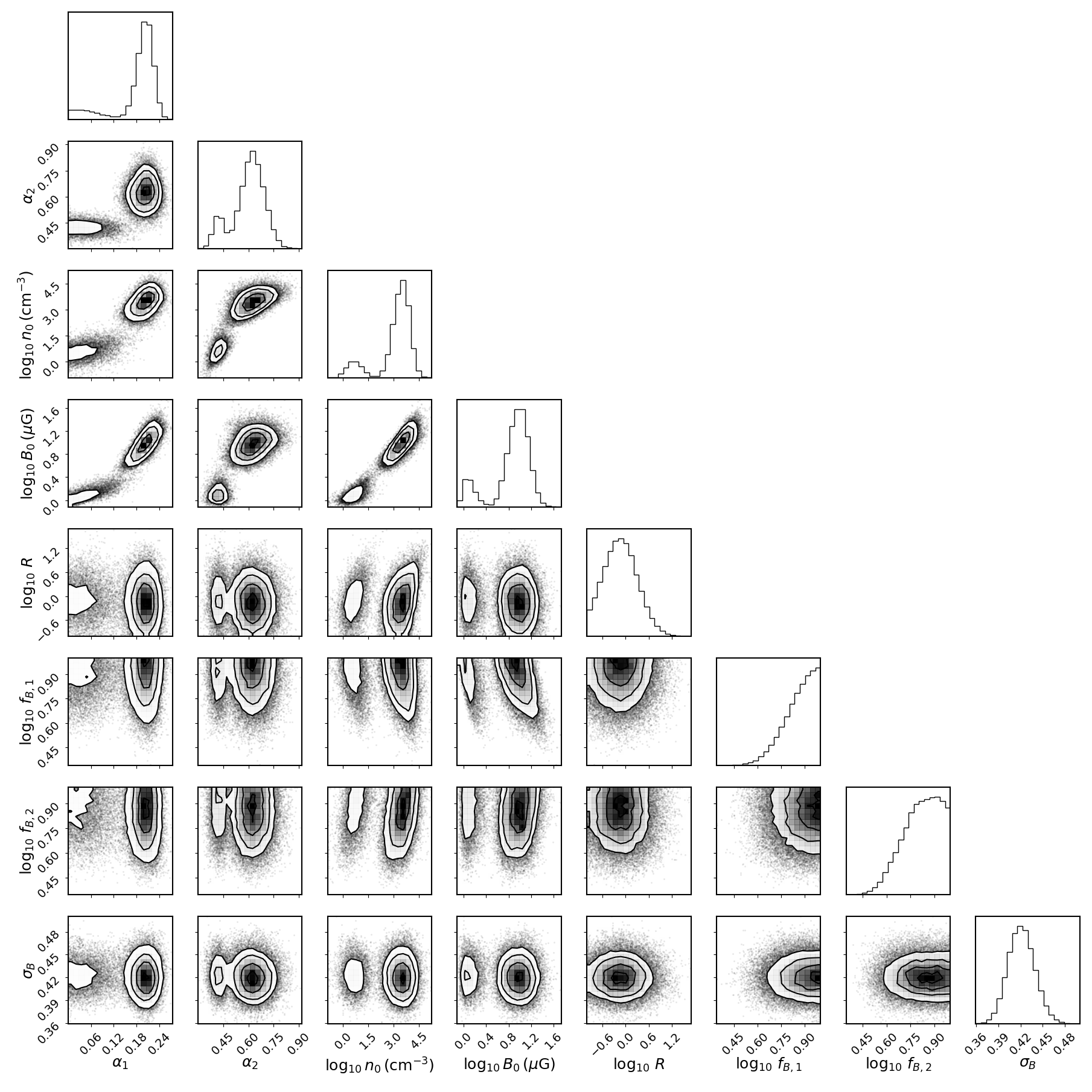}
    \caption{Corner plot for Model~D applied to DS6}
    \label{fig:corner_D_DS6}
\end{figure*}

The drawback of such an approach is immediately obvious since with fewer points in this part of the data set, the posterior is less well anchored and the break shifts. In addition, since $R_n$ controls the sharpness, or width of the transition, and is primarily constrained by data near the break, removing these points weakens the constraint on $R_n$, allowing it to vary more in response to the remaining data. This can cause $R_n$ to increase or decrease depending on the data set. The same loss of leverage also leads to slightly steeper slopes and a larger $B_0$. This change reflects not just the removal of low-field points, but the fact that those points disproportionately constrain the turnover in the relation. In addition, inspection of the corner plot, Figure~\ref{fig:corner_D_DS6}, reveals bimodality in the posterior, indicating degeneracy in the results. A notable feature is a secondary posterior concentration near $\alpha_1 \approx 0$. Although this mode has lower posterior support than the main solution, it indicates increased degeneracy in the diffuse-regime slope when Zeeman points below $3\,\mu$G are removed. In practice, the sensitivity cut reduces leverage in the transition region, allowing an approximately flat diffuse slope to remain statistically plausible. As this is only a diagnostic threshold test (not a full left-censored data treatment), we treat this feature cautiously and do not draw a strong physical conclusion from it. Combined with the sampling difficulties discussed above, this indicates that the models struggle to constrain the relation when the sensitivity cut is applied in this way. Therefore, we avoid using this approach in reporting our final results in the Conclusions.

\begin{table*}[!htbp]
    \begin{center}
    \caption{MAP values from sensitivity limit analysis of model~B on the reduced data sets}
    \begin{tabular}{lccccccccc}  
    \hline
    \\
        Data&  $\alpha_1$ & $\alpha_2$ & $n_0$ & $B_0 (\mu{\rm G})$ & $f_B$ & $\sigma_B$ & $R_n$ & $(f_n + 1)$ \\
        Set& &&$(10^3{\rm cm}^{-3})$ &&&&&&
        \\
        \\
        \hline \hline
        \\
        DS5 & $0.20_{-0.18}^{+0.01}$ & $0.62_{-0.06}^{+0.04}$ & $2.91_{-2.56}^{+0.65}$ & $85.64_{-80.26}^{+20.23}$ & $0.15_{-0.14}^{+0.30}$ & $0.15_{-0.12}^{+0.02}$ & 0.30 & 2 \\
        \\       
        DS6 & $0.21_{-0.04}^{+0.02}$ & $0.67_{-0.14}^{+0.06}$ & $2.88_{-2.88}^{+0.43}$ & $38.69_{-37.80}^{+6.72}$ & $0.26_{-0.25}^{+0.30}$ & $0.30_{-0.02}^{+0.04}$ & 0.30 & 2 \\
        \\
        \hline
        \\
        DS5 & $0.19_{-0.19}^{+0.09}$ & $0.62_{-0.23}^{+0.04}$ & $2.56_{-2.56}^{+0.12}$ & $66.06_{-66.05}^{+11.34}$ & $0.16_{-0.16}^{+0.27}$ & $0.08_{-0.08}^{+0.02}$ & 0.70 & 5 \\
        \\
        DS6 & $0.21_{-0.18}^{+0.01}$ & $0.65_{-0.37}^{+0.16}$ & $3.05_{-3.05}^{+2.99}$ & $31.25_{-31.24}^{+18.19}$ & $0.32_{-0.32}^{+0.21}$ & $0.05_{-0.05}^{+0.01}$ & 0.70 & 5 \\
        \\
        \hline
        \\
        DS5 & $0.24_{-0.24}^{+0.26}$ & $0.61_{-0.29}^{+0.05}$ & $3.86_{-3.86}^{+3.07}$ & $58.43_{-58.43}^{+24.44}$ & $0.22_{-0.22}^{+0.21}$ & $0.13_{-0.13}^{+0.01}$ & 0.95 & 9 \\
        \\
        DS6 & $0.11_{-0.11}^{+0.32}$ & $0.38_{-0.14}^{+0.02}$ & $0.001_{-0.001}^{+0.001}$ & $7.09_{-7.08}^{+3.00}$ & $0.16_{-0.16}^{+0.35}$ & $0.07_{-0.07}^{+0.00}$ & 0.95 & 9 \\
        \\
        \hline \hline    
   \end{tabular}
    \label{table:sensitivity_results_model_B}
    \end{center}
    \par\noindent\textit{Notes.} 
    The same headings and labels as Table \ref{table:model_b_results} except we have used the reduced data sets as described in the text above.
\end{table*}

As previously noted, a more rigorous treatment would explicitly incorporate the sensitivity limit into the likelihood, rather than excluding all sub-threshold values. In such a framework, measurements below the threshold would be treated as \textit{left-censored} data: non-detections would contribute to the likelihood through the probability that the true field strength lies below the sensitivity limit, while detections would contribute as usual. Numerically, this means that the reported value of $B$ below the sensitivity limit is treated as an upper limit, and the true value can lie between this reported value and zero. This \textit{left-censored} data modelling approach \citep{Loredo2004,Kelly2007} ensures that all information is used and that the inference accounts for observational selection effects imposed by the detection threshold. However, implementing such an extension to our approach would involve a much more detailed analysis and is beyond the scope of this work.

\begin{table*}[!htbp]
    \begin{center}
    \caption{MAP values from sensitivity limit analysis of model~C on the reduced data sets}
    \begin{tabular}{lccccccccc}  
    \hline
    \\
        Data &  $\alpha_1$ & $\alpha_2$ & $n_0$ & $B_0(\mu{\rm G})$ & $f_{B,1}$ & $f_{B,2}$ & $\sigma_B$ & $R_n$\\
        Set & & & $(10^3{\rm cm}^{-3})$ &&&&&
        \\
        \\
        \hline \hline
        \\       
        DS5 & $0.18_{-0.14}^{+0.03}$ & $0.64_{-0.08}^{+0.02}$ & $9.97_{-9.62}^{+5.12}$ & $18.34_{-12.54}^{+4.78}$ & $0.71_{-0.09}^{+0.29}$ & $0.88_{-0.10}^{+0.12}$ & $0.31_{-0.03}^{+0.03}$ & $1.34_{-1.29}^{+0.15}$ \\  
        \\
        DS6 & $0.21_{-0.03}^{+0.02}$ & $0.65_{-0.11}^{+0.06}$ & $4.83_{-4.83}^{+0.46}$ & $12.98_{-8.34}^{+0.67}$ & $0.80_{-0.03}^{+0.20}$ & $0.73_{-0.02}^{+0.26}$ & $0.41_{-0.01}^{+0.02}$ & $0.71_{-0.67}^{+0.47}$ \\
        \\
        \hline \hline
    \end{tabular}
    \label{table:sensitivity_results_model_C}
    \end{center}
    \par\noindent\textit{Notes.} 
    The same headings and labels as Table \ref{table:model_c_results} except we have used the reduced data sets as described in the text above.
\end{table*}
        
\begin{table*}[!htbp]
    \begin{center}
    \caption{MAP values from sensitivity limit analysis of model~D on the reduced data sets}
    \begin{tabular}{lccccccccc}  
    \hline
    \\
        Data &  $\alpha_1$ & $\alpha_2$ & $n_0$ & $B_0(\mu{\rm G})$ & $f_{B,1}$ & $f_{B,2}$ & $\sigma_B$ & $R_n$\\
        Set & & & $(10^3{\rm cm}^{-3})$ &&&&&
        \\
        \\
        \hline \hline
        \\
        DS6 & $0.22_{-0.04}^{+0.01}$ & $0.64_{-0.11}^{+0.07}$ & $3.56_{-3.56}^{+0.15}$ & $12.26_{-7.67}^{+1.63}$ & $0.86_{-0.03}^{+0.14}$ & $0.78_{-0.02}^{+0.22}$ & $0.42_{-0.01}^{+0.02}$ & $0.46_{-0.45}^{+0.34}$ \\
        \\
        \hline \hline
   \end{tabular}
    \label{table:sensitivity_results_model_D}
    \end{center}
    \par\noindent\textit{Notes.} 
    The same headings and labels as Table \ref{table:model_d_results} except we have used the reduced data sets as described in the text above.
\end{table*}

\section{Appendix D}
\label{app:single_slope}

Here we show the model set up, priors and results for testing out a single-slope power-law model based on our Model D and DS4. We have tried to keep the models as similar as possible in terms of priors, posterior sampling, and chains. However, due to this model now containing only one slope, we lose both $\alpha_2$ and $n_0$, as there is no break, as well as $f_{B,2}$.

The model therefore becomes
\begin{equation}
\label{single_power_law}
    |\mathbf{B}| = B_0 \times n^{\alpha},
\end{equation}
where $B_0$ is the normalisation parameter of the magnetic field, $n$ is the reported number density and $\alpha$ is the slope.

To implement the errors and log-likelihood we adopt a similar technique to those used in Models A and D such that the envelope scaling parameter transform in log-space is the same as Equation \ref{eq:b_transform}.

We also take Equation \ref{eq:b_transform} to calculate the effective error $n_{\mathrm{eff},i}$ on $n$.

The priors we used for this set up are taken from Model D and shown in Table \ref{table:single_slope_priors}. We also use the same chain lengths of $5\,000$ warm-up iterations followed by $5\,000$ sampling iterations and the same \texttt{adapt\_delta} $=0.9$ and \texttt{max\_treedepth} $=12$.

\begin{figure}[!ht]
    \centering
    \includegraphics[scale=0.35]{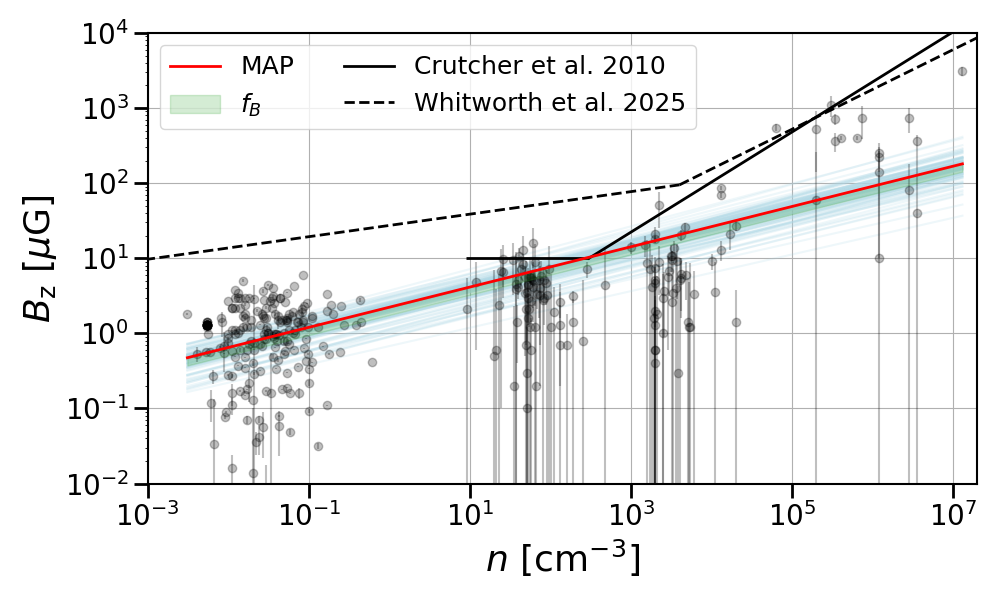}
    \caption{MAP result for the single slope model based on Model D and DS4.}
    \label{fig:single_slope}
\end{figure}

Table \ref{table:single_slope} shows the results for this model, while Figure \ref{fig:single_slope} shows the MAP result for the single slope model. We can see here that, visually, a single slope model could be considered viable, but it fails to take into account the change in physical regimes that occurs between the diffuse and dense gas in the ISM, which a broken power-law does.

\begin{table}
\begin{center}
\caption{Model hyperparameters and priors}
\begin{tabular}{llccc}
\hline \\
 & Parameter & Single Slope \\
\hline \hline
\\
$B_0$ [$\log_{10} B_0$] & Envelope normalization & 
  $\log_{10} B_0 \sim \mathcal{N}(0,2)$ \\ 
 & ($\mu$G) & $>-10-Y_\mathrm{min}$ \\ 
\\
$\alpha$ & $n$ slope & $\mathcal{N}(0.6,0.2)$ \\ 
 & & $[0,1]$ \\ 
\\
$f_B$ [$\log_{10} f_B$] & Envelope scaling & $\log_{10} f_B \sim \mathcal{N}(0,0.1)$ \\ 
 & & $[10^{-4},1]$ \\ 
\\
$\sigma_B$ & $B$ intrinsic scatter & $\mathcal{N}(0.1,0.1)$  \\ 
 & & $[0,10]$ \\ 
\\
$f_n$ [$\log_{10} f_n$] & intrinsic scatter  
& $\log_{10} f_n \sim \mathcal{N}(0,1.0)$ \\
 & & $[10^{-2},10^3]$ \\ 
\\
\hline \hline
\label{table:single_slope_priors}
\end{tabular}
\end{center}
\par\noindent
\textit{Notes.} The top line for each parameter shows the prior choice whilst the bottom shows the HMC bounds.
\end{table}

\begin{table}[!htbp]
    \begin{center}
    \caption{MAP values from the single-slope analysis of model~D on the Zeeman plus Pulsar data set with Pulsar derived errors on $n$}
    \begin{tabular}{lccccc}  
    \hline
    \\
        Data & $\alpha$ & $B_0(\mu{\rm G})$ & $f_B$ & $\sigma_B$ & $R_n$\\
        Set & & & & &
        \\
        \\
        \hline \hline
        \\
        DS4 & $0.27_{-0.01}^{+0.01}$ & $0.48_{-0.17}^{+0.09}$ & $0.79_{-0.01}^{+0.21}$ & $0.45_{-0.02}^{+0.02}$ & $1.01_{-1.00}^{+1.72}$ \\
        \\
        \hline \hline
   \end{tabular}
    \label{table:single_slope}
    \end{center}
    \par\noindent\textit{Notes.} 
    The same headings and labels as Table \ref{table:model_d_results} except we have used the reduced data sets as described in the text above.
\end{table}

\end{appendix}

\end{document}